\documentclass[%
 reprint,
superscriptaddress,
amsmath,amssymb,aps,
pra,notitlepage,longbibliography
]{revtex4-1}

\usepackage{natbib}
\usepackage{braket}
\usepackage[pdftex]{graphicx}
\usepackage{mathrsfs}
\usepackage{verbatim}
\usepackage[toc,page]{appendix}
\usepackage{comment}
\usepackage{xr}
\usepackage{xcolor}
\usepackage{longtable}

\usepackage[breaklinks,colorlinks=false,linkcolor=blue,urlcolor=blue,citecolor=red]{hyperref}

\begin{document}

\title{Emergent non-Fermi liquid phenomena in multipolar quantum impurity systems}
\author{Adarsh S. Patri}
\affiliation{Department of Physics and Centre for Quantum Materials, University of Toronto, Toronto, Ontario M5S 1A7, Canada}

\author{Ilia Khait}
\affiliation{Department of Physics and Centre for Quantum Materials, University of Toronto, Toronto, Ontario M5S 1A7, Canada}

\author{Yong Baek Kim}
\affiliation{Department of Physics and Centre for Quantum Materials, University of Toronto, Toronto, Ontario M5S 1A7, Canada}


\begin{abstract}

Discovery of novel spin-orbital entangled quantum ground states paves an important avenue for controllable quantum materials via unique couplings to the lattice and other external perturbations. In this work, motivated by recent experiments on cubic heavy fermion materials with multipolar local moments, we theoretically investigate strongly-interacting spin-orbital entangled quantum ground states in multipolar quantum impurity systems. Here itinerant electrons are interacting with the local moments carrying quadrupolar and octupolar moments, in contrast to the conventional Kondo problem with dipolar local moment. Using perturbative renormalization group methods, we uncover a number of non-Fermi liquid ground states, which are characterized by an absence of well-defined quasiparticles and singular power-law behaviours in physical properties. We show that the non-Fermi liquid states found here are outside the known categories of non-Fermi liquid states in the conventional multi-channel Kondo problem. This work lays a novel ground for the identification of unexpected non-Fermi liquid phases in many strongly spin-orbital-coupled quantum materials.

\end{abstract}

\maketitle

\section{Introduction}

Understanding non-Fermi liquid states may be the key to explaining unconventional
superconductivity and other broken symmetry states in strongly-correlated itinerant electron
systems\cite{Keimer_cuprates,pnictide_si, chubukov_fe,Coleman_perspective,Hewson_1993,Phillips}.
This is in contrast to the Fermi liquid states with well-defined quasiparticles, which
is the basis for our understanding of weakly-correlated electron systems. 
In the traditional picture of heavy fermion materials, the RKKY interaction~ \cite{RKKY}
between dipolar local moments (or spins) competes with the formation of a Fermi liquid
state with a large Fermi surface \cite{Doniach_1977,Khait}, facilitated by the Kondo coupling between the dipolar
local moment and conduction electrons. This competition leads to the emergence of non-Fermi liquid {behaviours}, which (as seen in various experiments \cite{SSLee,Stewart}) are often attributed to quantum critical phenomena associated with such quantum phase transitions.

Such a simple picture, however, may change drastically if the local moments are higher-order 
multipolar moments (and not merely dipolar) \cite{multupole_rev_1, review_exotic_multipolar}. 
This is because the standard picture of the screened dipolar local moments 
by conduction {electrons} may not be applicable to the multipolar cases.
In fact, local moments in many heavy fermion systems carry multipolar moments
because of strong spin-orbit coupling and crystal electric field effects \cite{hfm_superconductivity_v, pr_fq, pr_afq, super_fq_pr_ti, pressure_hfm_super_quad_pr_ti, Fisher_2015, fisher_2019, sbl_ybk_ap_2018, hattori_afq_fq_2014, sbl_ybk_landau_2018, patri_unveil_2019}.
The fate of such an unusual Kondo effect would also have ramifications in
multipolar ordering induced by the same coupling, often dubbed ``hidden order'' 
due to the difficulty of detecting it with conventional experimental probes \cite{hidden_order_review, multupole_rev_3}.
An earlier investigation of this issue \cite{cox_quad_kondo_1987} suggests that when the quadrupolar moment
of a single local impurity couples to conduction electrons, the resulting Kondo effect
leads to a non-Fermi liquid state. Here the orbital degrees of freedom of the 
conduction electrons are entangled with the fluctuations of quadrupolar moments
while the spin quantum number of conduction electrons merely provides 
two separate channels for the scattering \cite{cox_quad_kondo_hfm, cox_quad_kondo_1987}. This model has been used
for further exploration of non-Fermi liquid physics.

In this paper, we investigate multipolar quantum impurity systems, where the local 
moment carries both quadrupolar and octupolar moments, and are coupled
to conduction electrons.
We are motivated by a series of experiments \cite{hfm_superconductivity_v,pr_nfl, quad_nfl_onimaru, pr_v_nfl} performed on 
the cubic heavy fermion systems, Pr(TM)$_2$Al$_{20}$ (TM = Ti, V), 
where the non-Kramers doublet of Pr$^{3+}$ ions carry
quadrupolar and octupolar moments. Here we consider
a single multipolar impurity coupled to conduction electrons in
generic symmetry-allowed orbital channels.
Using perturbative renormalization group analysis, we establish
the presence of more than one possible non-Fermi liquid state,
signified by the presence of non-trivial fixed points in the renormalization group flow.
These include a novel non-Fermi liquid state, which was not identified
in previous studies, while the others are connected to the
two-channel Kondo effect mentioned above.
The newly discovered non-Fermi liquid state is characterized by 
the absence of well-defined quasiparticles, and power-law 
behaviours in response functions that are more singular than the two-channel Kondo counterpart. 
In this case, quantum fluctuations of the multipolar moments are 
highly entangled with both the orbital and spin degrees of freedom of conduction electrons. This is in contrast to the two-channel Kondo model, where only the orbital is entangled with local moment fluctuations, while spin remains a spectator.
{Indeed, the discovered novel non-Fermi liquid is outside the paradigm of the multi-channel Kondo effect.}
We expect that our results would serve as a valuable starting point towards understanding
non-Fermi liquid states, quantum phase transitions, and critical phenomena 
associated with ``hidden order'' systems.

\section{Models}

{
We first consider the origin and nature of the local moments that arise in the archetypal multipolar heavy fermion system, as well as the source of the conduction electrons.
We then present the subsequent `Kondo' Hamiltonians, highlighting the importance of the orbital dependence on the form of the Kondo coupling.
In this aspect, we start by writing an effective (low-energy) `Kondo' Hamiltonian from pure symmetry considerations, without an emphasis on any particular microscopic model.
This approach is attractive as it ensures that we are not restricted to any particular microscopic model to begin with, nor does it require us to precisely know the microscopic processes and mechanisms responsible for the low-energy interaction.
As such, we are able to explore (symmetry-permitted) generic forms of Kondo couplings that can arise in this system.
Nevertheless, to place our work in the context of previous quadrupolar Kondo models \cite{cox_quad_kondo_hfm, cox_quad_kondo_1987}, we also present a microscopic inter-site Anderson model, which involves the hybridization of conduction electrons with the local moment $f$-electron states.
Indeed, in the low-energy limit, this Anderson model reduces to our effective `Kondo' models, thus providing an explicit microscopic origin of our low-energy theory.
}

\subsection{Local moment physics of P\lowercase{r} ions} \label{sec_micro_local}

An ideal representative example of multipolar heavy fermion systems is the cubic rare-earth family Pr(TM)$_2$Al$_{20}$ (TM = Ti, V), where the localized multipolar degrees of freedom arise from Pr 4$f^2$ electrons. In these compounds, each Pr ion is surrounded by a Frank-Kasper (FK) cage of Al atoms (16 Al-atom polyhedra), which subjects the $f$ electrons to a crystalline electric field (CEF) of local $T_d$ symmetry. This CEF splits the $J=4$ multiplet of the 4$f^2$ electrons to yield ground states formed by a $\Gamma_{3g}$ non-Kramers doublet. The $\Gamma_{3g}$ states support time-reversal even quadrupolar moments $\mathcal{O}_{20} = \frac{1}{2} (3J_z^2 - J^2)$, $\mathcal{O}_{22} = \frac{\sqrt{3}}{2} (J_x^2 - J_y^2)$, and time-reversal odd octupolar moment $\mathcal{T}_{xyz} = \frac{\sqrt{15}}{6} \overline{J_x J_y J_z}$, where the overline represents a fully symmetrized product. Using a pseudospin basis $ \{ \ket{\uparrow}, \ket{\downarrow} \} $ from the doublet (as described in Appendix A), we can write the multipolar moments in terms of an effective pseudospin-1/2 operator $\vec{S} = \left( S^x, S^y, S^z \right)$,
\begin{align}
&S^x = -\frac{1}{4}\mathcal{O}_{22},  ~~~ ~~~S^y = -\frac{1}{4}\mathcal{O}_{20},  ~~~ ~~~S^z = \frac{1}{3\sqrt{5}}\mathcal{T}_{xyz}. 
\end{align}
The XY components of the pseudospin vector represent the quadrupolar degrees of freedom, while the Ising direction describes the octupolar degree of freedom.

\subsection{Conduction electron Model} \label{sec_micro_conduction}

We now turn to the conduction electrons which, in this family, {arise primarily} from the Al atoms and the TM ions. Since our principal interest in this work is on the Kondo coupling of conduction electrons to localized Pr moments, we focus on the itinerant electrons emanating from the Al atoms, and ignore the contributions from the TM ions. This choice stems
from the physically justifiable expectation that cage compounds well isolate the (Pr) atom located at the centre of the cage from the rest of its surroundings.
We now focus on a single FK cage of Al atoms. Due to the cubic symmetry of the FK cage, the eigenstates of the electrons hopping on a single cage can be organized into the irreducible representations of the local $T_d$ group. Importantly, each eigenmode $A_1, T_2, E,...$ has the same symmetry character as $s, p, e_g,...$ electron orbitals centred about the FK cage. We can thus interpret the eigenmodes as being equivalent to electron \textit{molecular} orbitals that are located about the centre of the cage, and the conduction Hamiltonian describes hopping between these molecular orbitals on different diamond sites.
Equipped with these molecular orbitals, there is still, however, an ambiguity as to which orbitals contribute to the Fermi surface and thus {to the form of the Kondo coupling}. Recent de Haas-van Alphen (dHvA) studies  \cite{dhva_nagashima_2013} have suggested a localized Fermi pocket about the zone centre for PrTi$_{2}$Al$_{20}$, although the orbital character of {the Fermi surface} is still unknown. We can nevertheless proceed by selecting conduction electrons in $e_g$-like and $p$-like molecular orbitals. This choice is plausible as the subsequently obtained tight-binding band structure recovers a well-localized Fermi surface about the zone centre as described in Appendix B.
More importantly, this selection of molecular orbitals is the simplest choice to obtain non-trivial Kondo couplings with the localized moments: $e_g$ orbitals yield the familiar two-channel Kondo model, while $p$ orbitals result in novel Kondo couplings that permit both conduction electron spin and orbital to participate in the quantum scattering events.
\textcolor{black}{For completeness, we note that the partially filled nature of the TM ions' $d$-orbitals do not contribute to the below described `Kondo' physics, as experimental resistivity \cite{pr_nfl} and dHvA \cite{dhva_nagashima_2013} studies on the isostructural La(Ti,V)$_{2}$Al$_{20}$ indicate that the responsibility lies with the $f$-electron moment arising from the Pr ions.}

\subsection{$e_g$ orbital Kondo model}

We first consider conduction electrons residing in the $e_g$ orbitals. The form of the Kondo coupling to the multipolar moments is constrained by the local $T_d$ symmetry imposed by the FK cage; we detail in Appendix C the manner in which the multipolar moments and the conduction electron degrees of freedom transform under the generating elements of $T_d$. The symmetry-permitted coupling is of the form,
\begin{align}
&{H}_Q^{e_g} = J_Q   c_{j,a,\alpha}^{\dagger} \left[ S_j^x  \sigma^0_{\alpha \beta} \otimes \tau^x_{ab}  -  S^y_j   \sigma^0_{\alpha \beta} \otimes \tau^z_{ab}  \right]  c_{j,b,\beta}  \ ,
 \nonumber \\
&{H}_O^{e_g} =  - J_O   S^z_j  c_{j,a,\alpha}^{\dagger} \left[  \sigma^0_{\alpha \beta} \otimes \tau^y_{ab} \right]  c_{j,b,\beta} ,
\label{eq_eg_Kondo}
\end{align}
where repeated Latin (Greek) indices are implicit summations over orbital (spin) degrees of freedom, and $\tau$ is the usual spin-1/2 Pauli matrix describing the $e_g$ orbital degree of freedom $ \left\{ x^2-y^2, \ 2z^2 - x^2 - y^2 \right \}$ as listed in Appendix D. We note that $j$ is the site where the (impurity) multipolar moment resides. The $e_g$ orbitals introduce two Kondo couplings: quadrupolar $J_Q$ and octupolar $J_O$. We note that the coupling in Eq. \ref{eq_eg_Kondo} has precisely the same form as the two-channel Kondo model, where here the conduction electron spin plays the role of a channel index. Although the sign structure in Eq. \ref{eq_eg_Kondo} may seem dissimilar to the usual two-channel Kondo model, this is merely an artifact of the choice of multipolar basis; a unitary transformation of the multipolar moments' basis allows the familiar form to be recovered.

\subsection{$p$ orbital Kondo model}

We next turn to the problem of coupling $p$ orbital conduction electrons to the local moments. 
The symmetry allowed couplings are of the form,
\begin{widetext}
\begin{align}
& H_{Q1}^{p} = K_{Q1} c_{j,a,\alpha}^{\dagger}   \left[ S^x_j  \sigma^0_{\alpha \beta} \otimes  \lambda^{x^2 - y^2}_{ab}  -  {S^y_j} \sigma^0_{\alpha \beta} \otimes  \lambda^{2z^2 - x^2 -y^2}_{ab} \right]  c_{j,b,\beta}, \nonumber \\
&H_O^{p} = K_O S^z_j  c_{j,a,\alpha}^{\dagger} \left[ \sigma^x_{\alpha \beta} \otimes \lambda^{yz,r}_{ab} + \sigma^y_{\alpha \beta} \otimes \lambda^{xz,r}_{ab}+ \sigma^z_{\alpha \beta} \otimes \lambda^{xy,r}_{ab} \right]  c_{j,b,\beta}, \nonumber \\
& {H}_{Q2}^{p} = K_{Q2} c_{j,a,\alpha}^{\dagger} \left[ \left( \sqrt{3}S^x_j - S^y_j \right)  \sigma^x_{\alpha \beta} \otimes \lambda^{yz,i}_{ab} +  \left( \sqrt{3}S^x_j + S^y_j \right) \sigma^y_{\alpha \beta} \otimes \lambda^{xz,i}_{ab}   +  2 S^y_j    \sigma^z_{\alpha \beta} \otimes \lambda^{xy,i}_{ab} \right]  c_{j,b,\beta},  
\label{eq_h_o_1}
\end{align}
\end{widetext}
where repeated Latin (Greek) indices are implicit summations over orbital (spin) degrees of freedom, and $\lambda$ are the SU(3) Gell-Mann matrices describing the orbital degree of freedom $ \left\{ p_x, p_y, p_z \right \}$ as listed in Appendix D. We note that $j$ is the site where the (impurity) multipolar moment resides. 
{The inter-orbital hopping $\lambda^{ab}$ matrices include the superscript of $r$ and $i$, indicating whether it contains solely $r$eal or $i$maginary matrix elements. This notation allows one to easily notice that the terms in Eq. \ref{eq_h_o_1} preserve time-reversal symmetry; the $\lambda^{ab}$ matrices containing $i$-superscript flip sign, and this sign change is compensated by the sign flip of the spin-1/2 Pauli matrices, $\vec{\sigma}$.}
The $p$ orbitals introduce three Kondo couplings: quadrupolar $K_{Q1}, K_{Q2}$ and octupolar $K_O$.  Both of $H_{Q1}^{p}$ and $H_{O}^{p}$ can be simply understood as the conduction electrons imitating the respective multipolar moments they are interacting with, in an effort to preserve the local $T_d$ symmetry. For instance, $H_{Q1}^{p}$ describes the scattering phenomena of conduction electrons distorting their electron charge distribution to a quadrupolar form of $x^2 - y^2$ and $2z^2 - x^2 - y^2$ to interact with the time-reversal even quadrupolar moments $J_x^2 - J_y^2$ and $2J_z^2 - J_x^2 - J_y^2$, respectively. The spin degree of freedom remains as a spectator here. Similarly, $H_{O}^{p}$ describes the phenomena of conduction electrons forming an $xyz$-like octupolar moment from its spin and orbital degrees of freedom to interact with the time-reversal odd octupolar moment $J_xJ_yJ_z$. This simple picture is, however, difficult to apply to ${H}_{Q2}^{p}$ as it involves the quadrupolar moments interacting with conduction electrons that undergo flip of both spin and orbital. One can minimally say that this term amplifies the quantum nature of the scattering event between the conduction electrons and local moment, as it involves both spin and orbital degrees of freedom.

\subsection{Combination of all orbitals: inclusion of $e_g \otimes p$ Kondo model}

Finally, we consider Kondo coupling involving the mixing of $e_g$ and $p$ orbitals. 
The actual form of the symmetry-constrained Hamiltonian is, however, not very illuminating, so it suffices here to state that we have three additional Kondo couplings: two involving interactions with the quadrupolar moments ($L_{Q1}, L_{Q2}$) and one involving the octupolar moment ($L_{O}$). This brings us to a grand total of eight coupling constants in the complete model: $J_Q, J_O$ from $e_g$, $K_{Q1}, K_{Q2}, K_O$ from $p$, and $L_{Q1}, L_{Q2}, L_O$ from $e_g \otimes p$. 
We present the complete Hamiltonian in Appendix I.

\subsection{Microscopic origin of Kondo couplings from an inter-site Anderson Model}
{
In considering an inter-site Anderson model, it is necessary to consider the energy level scheme of the $f$ electrons.
The Pr$^{3+}$ ion can fluctuate from its 4$f^2$ $\Gamma_3$ non-Kramers ground state to Kramers 4$f^1$ and 4$f^3$ excited states via hybridization with a bath of conduction electrons.
For the purpose of this discussion, it suffices to consider the $f^1$ doublet excited levels, which is of $\Gamma_7$ symmetry (the $f^3$ state is assumed to be at much higher energy than the $f^1$ state, and therefore ignored \cite{cox_quad_kondo_hfm, cox_quad_kondo_1987}). 
The basis states of the $f^1$ level are presented in Appendix A.
Group theoretically, a valence fluctuation from the $f^2$ ($\Gamma_3$) to $f^1$ ($\Gamma_7$) levels is only permitted by conduction electrons with specific symmetry: $\Gamma_c = \Gamma_3 \otimes \Gamma_7 = \Gamma_8$, where $\Gamma_c$ is the conduction electron irrep.
In Ref.~\onlinecite{cox_quad_kondo_hfm, cox_quad_kondo_1987}, the hybridizing conduction electrons were taken to be a $\Gamma_8$ quartet of partial waves, which were constructed from $l=3$ angular momentum states, which were spin-orbit coupled to produce $j=\frac{5}{2}, \frac{7}{2}$ manifolds.
This analysis led to the storied two-channel Kondo model at low-energy.
Instead, we can consider other means by which to construct the $\Gamma_8$ irreducible representation. 
One possibility is that of $\Gamma_c = \Gamma_8 = e_g \otimes \frac{1}{2}$, where the $e_g$ orbitals transform as $\Gamma_3$ and $\frac{1}{2}$ is the spin-1/2 spinor states that transform as $\Gamma_6$. 
Such conduction orbitals, as will be shown, is the microscopic origin of the above $e_g$ Kondo model.
Another possibility is that of $\Gamma_c = p \otimes \frac{1}{2} = \Gamma_8 \oplus \Gamma_6$, which can also mediate the $f^2 \leftrightarrow f^1$ transition; here, $p$ transforms as a $\Gamma_4$ irrep. 
Thus, group theoretically, conduction $e_g$ and $p$ orbitals equipped with spinor degrees of freedom are permitted to hybridize with the Pr$^{3+}$ energy levels. 
}

{
We now present an explicit demonstration of the above group theory discussion by considering a single four-site tetrahedron cage of Al atoms, each hosting atomic $p$ orbitals.
The choice of atomic orbitals is justified from the valence electronic configuration of an Al atom [Ne]3s$^2$3p$^1$.
The tetrahedron geometry is indeed a simplification of the more complicated FK cage, but is valid as it still imposes the constraining $T_d$ symmetry of the cage.
Considering a tight-binding hopping matrix on this single tetrahedron, the corresponding molecular orbitals (eigenstates) are decoupled as: $A_1 \oplus E \oplus T_1 \oplus 2T_2$, where we use the $T_d$ irrep notation. Here $E$ and $T_2$ have $e_g$ and $p$ as its basis functions. 
In abiding with the $T_d$ symmetry, we can derive the following inter-site Anderson model,
\begin{align}
H_{\rm hyb} = \sum_{\Delta, \omega, j, \mathcal{P}, \sigma} V_{\Delta  \omega j \mathcal{P}  \sigma} \ket{f^2, \Delta} \bra{f^1, \omega} c_{j \mathcal{P} \sigma} + \rm{h.c.}
\end{align}
where $V_{\Delta  \omega j \mathcal{P}  \sigma}$ is the hybridization matrix element, $\Delta=\{1,2 \}$ sums over the two non-Kramers $\Gamma_3$ ground states, $\omega = \{ +,- \} $ runs over the two $f^1$ levels, $j=\{ 1,2,3,4 \}$ goes over the four-sites, $\mathcal{P}= \{p_x, p_y, p_z \}$ runs over the atomic Al $p$ orbitals, and $\sigma = \{ \uparrow, \downarrow \}$ denotes the conduction electron spin.
Constrained by symmetry, the ninety-six hybridization elements are composed of four independent parameters; we present its relations in Appendix E.
Rewriting the above Anderson model in terms of the molecular orbitals on the cage, and performing a Schrieffer-Wolff transformation, as described in Appendix F, we thus obtain: (i) for the $e_g$ molecular orbitals, the isotropic two-channel Kondo model of Eq. \ref{eq_eg_Kondo}, and (ii) the $p$ orbital Kondo model of Eq. \ref{eq_h_o_1}.
As we show next, starting with any of these terms would generate all the symmetry allowed interactions during the renormalization group (RG) procedure. }

%
%
%
%

\begin{figure*}[t]
\includegraphics[scale=1.0]{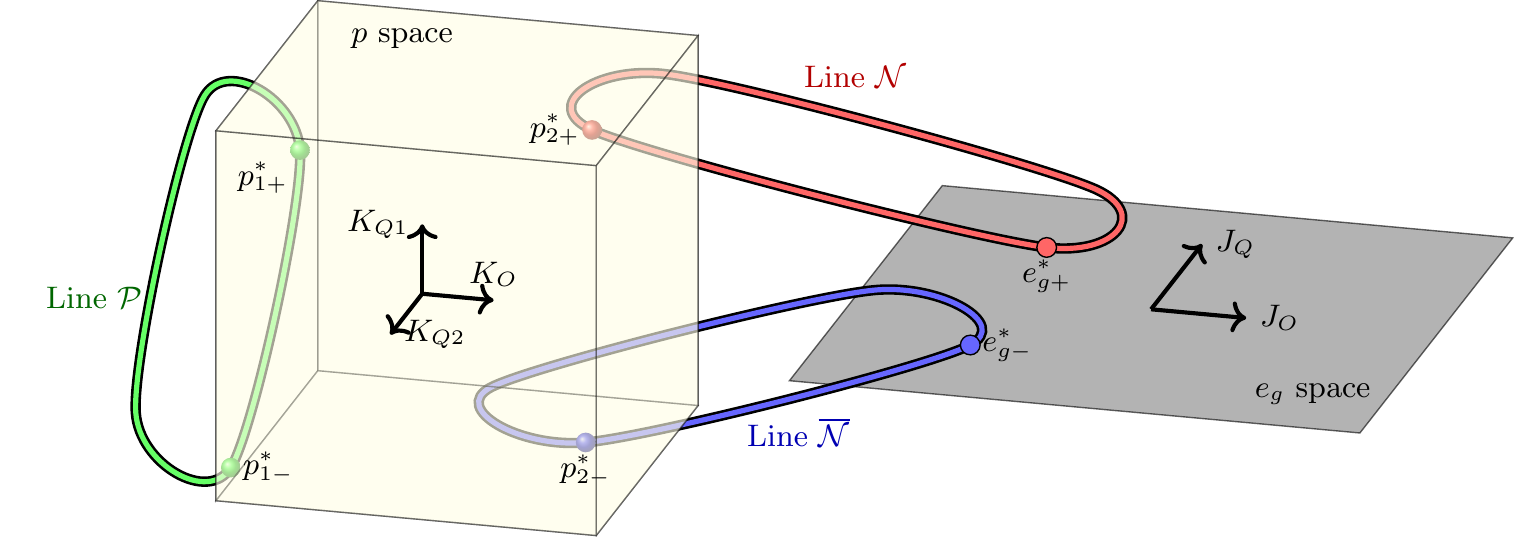}
\caption{Schematic plot denoting RG fixed lines and points in eight dimensional parameter space of the complete multipolar quantum impurity model. The `yellow-shaded cube' denotes the $p$-model space, with its corresponding fixed points at $ p_{1\mp}^*$ and $ p_{2 \mp}^*$ indicated by `green' spheres, and `red' and `blue' spheres respectively. The $e_g$ space is depicted by the grey-shaded plane, which includes the Nozi\`eres fixed points $e_{g\pm}^*$. The connective Lines  $\mathcal{N}$, $\overline{\mathcal{N}}$, $\mathcal{P}$ are denoted by red, blue and green one-dimensional curves, respectively, which reside in the three extra dimensions (that are not shown in the figure).}
\label{fig_region_all}
\end{figure*}

\section{Results}
{
Equipped with the low-energy models of the previous section, we examine the RG fixed points of our models by performing perturbative RG analysis of the vertex function as described in Appendix G.
We present the RG equations and delineate the fixed points/lines of the $e_g$ and $p$ orbital models, before presenting a schematic plot of the parameter space, and the location of the various fixed points/lines.
Finally, we discuss the physical properties associated with each of these fixed points.
}
\subsection{Renormalization group analysis of $e_g$ model: two channel Kondo model}
The perturbative renormalization group analysis of Eq. \ref{eq_eg_Kondo} (i.e. the two-channel Kondo model) yields the following $\beta$ functions for the two coupling constants,
\begin{align}
\frac{d J_{Q}}{d \ln D} &= 2 J_Q J_O  +  2J_{Q} \left(  J_Q ^2 + J_O^2 \right), \nonumber \\
\frac{d J_{O}}{d \ln D} &= 2 J_Q^2 +  4J_{O}  J_{Q}^2,   
\end{align}
whose fixed points are the decoupled Gaussian fixed point $G^* \equiv (J_Q=J_O=0$), and non-trivial fixed points $e_{g\pm}^* \equiv (J_Q, J_O) = (\pm 1/2, -1/2)$. $e_{g\pm}^*$ are known as the Nozi\`eres fixed points, which were found in perturbative RG computations of the two-channel Kondo model \cite{noz_fixed_point}. We also obtain a line of fixed points characterized by $J_Q = 0, J_O \neq 0$. The Gaussian fixed point is saddle-like (one relevant and one irrelevant eigendirections), while the fixed line has regions where the nearby flow is attractive ($J_O>0$) and repulsive ($J_O<0$). We present the RG flow diagram in Appendix H. The non-trivial fixed points are related to each other by the transformation of $S^{x,y} \rightarrow -S^{x,y}$; this transformation preserves the canonical commutation rules for the pseudospin-1/2 operators. The $\beta$ function and the characteristics of the fixed points/lines are precisely the same as the usual two-channel Kondo model. This leads to the same physical behaviour and exponents, as will be discussed.

\subsection{Renormalization group analysis of $p$ orbital Kondo model}

The perturbative renormalization group analysis of Eq. \ref{eq_h_o_1} yields the following $\beta$ functions for the three coupling constants,
\begin{align}
\frac{d K_{Q1}}{d \ln D} &= 6 K_{Q2} K_{O} +  K_{Q1} \left( 12 K_{Q2}^2 + 6 K_{O}^2 + 2 K_{Q1}^2 \right), \nonumber \\  
\frac{d K_{Q2}}{d \ln D} &= K_{O} \left(K_{Q1} - \sqrt{3} K_{Q2} \right) +  K_{Q2} \left( 12 K_{Q2}^2 + 6 K_{O}^2 + 2 K_{Q1}^2 \right),  \nonumber \\  
\frac{d K_{O}}{d \ln D} &= 4 K_{Q1} K_{Q2}  - 2 \sqrt{3} K_{Q2}^2 +  K_{O} \left( 24 K_{Q2}^2 + 4 K_{Q1}^2  \right),  
\end{align}
whose fixed points are [in terms of $(K_{Q1}, K_{Q2}, K_O)$]  the Gaussian fixed point $G^* = (0,0,0)$, $p_{2 \mp}^*= (\mp \frac{1}{2 \sqrt{3}},\pm \frac{1}{6}, \frac{1}{2 \sqrt{3}})$, and $p_{1 \mp}^* = (\mp \frac{1}{2 \sqrt{6}} , \mp \frac{1}{12\sqrt{2}},  - \frac{1}{4 \sqrt{3}})$. The non-trivial fixed points are stable and attractive. We present a representative RG flow diagram in Appendix H, where we depict the flow about the $p_{1 \mp}^*$ and $p_{2 \mp}^*$ fixed points. We also obtain a fixed line characterized by $K_O \neq 0$ ($K_{Q1} = K_{Q2} = 0$), which is analogous to the fixed line from the usual spin-Kondo model. Just as in that instance, it describes an Ising-like interaction between the conduction electron and local moment. The subtlety here is that the Ising degree of freedom is octupolar, as opposed to dipolar in the usual Kondo problem.

\subsection{Renormalization group analysis of $e_g \otimes p$ orbital Kondo model}

{For brevity}, we present the complete corresponding eight RG flow equations in Appendix I, and also include the complete $\beta$ functions for the $e_g$ and $p$-only couplings, which have been modified by the inclusion of these three additional couplings.
The solutions to the eight $\beta$ functions provide three extended lines of stable fixed points, two unstable fixed points, and an Ising plane of fixed points. The Ising plane is a generalization of the Ising `line' of fixed points of the $e_g$ and $p$ only models; specifically, all the quadrupolar coupling constants are zero, while the octupolar couplings are non-vanishing. We focus on the three stable (from the perspective of linear stability analysis) lines of interest, each of which are parametrized by a single coupling parameter. We present in Fig. \ref{fig_region_all} an illustrative schematic of the stable lines of fixed points to help conceptualize their relation to each other in the eight-dimensional parameter space. 
Importantly, the three connective lines of fixed points depicted by Lines $\mathcal{N}$ (`red'), $\overline{\mathcal{N}}$ (`blue'), and $\mathcal{P}$ (`green') are described by a single coupling parameter and include (depending on the line of interest) some of the $e_g$-only and $p$-only fixed points:

\noindent (i) Line $\mathcal{N}$ -- denoted as a `red' curve in Fig. \ref{fig_region_all} -- is parametrized by single coupling parameter, $J_Q \in [0, 1/2] $ (density of $e_g$ orbitals coupling to the quadrupolar moments), and includes the Nozi\`eres $e_{g+}^*$ fixed point, and a $p$ orbitals-only fixed point $p_{2+}^*$. 

\noindent (ii) Line $\overline{\mathcal{N}}$ -- denoted as a `blue' curve in Fig. \ref{fig_region_all} -- is parametrized by single coupling parameter, $J_Q \in [-1/2, 0]$ (density of $e_g$ orbitals coupling to the quadrupolar moments), and includes the related-to-Line $\mathcal{N}$ $e_{g-}^*$ and $p_{2-}^*$ orbitals-only fixed points. Line $\overline{\mathcal{N}}$ is mapped to Line $\mathcal{N}$ by taking $J_Q \rightarrow -J_Q$.

\noindent (iii) Line $\mathcal{P}$ -- denoted as a `green' curve in Fig. \ref{fig_region_all} -- is parametrized by single coupling parameter, $K_{Q1} \in [-\frac{1}{2\sqrt{6}}, \frac{1}{2\sqrt{6}}]$ (density of $p$ orbitals coupling to the quadrupolar moments), and includes two (related to each other) distinct $p$ orbitals-only fixed points $p_{1\mp}^*$. These $p$-orbitals-only fixed points are distinct from the ones located in Lines $\mathcal{N}$ and $\overline{\mathcal{N}}$.

\subsection{Physical Properties: Specific Heat, Resistivity}

The nature of the fixed points (lines) can be better understood by examining their influence on the physical properties of the conduction electrons, namely specific heat $c_v$ and electrical resistivity $\rho$. Since each of the isolated fixed points of the $e_g$-only and $p$-only models are included in the connective lines, it suffices to discuss the physical properties of each of the stable lines. We follow the perturbative approach used in Ref. \onlinecite{Gan_1994} to compute these two observables, and describe in Appendix J, K, L 
the schemes used in obtaining the exponents. We present a summary of the scaling behaviour of the specific heat and resistivity in the three connective lines in Table \ref{tab_phys_all}. We note that the total scattering/decay rate ($1 / \tau_{tot}$) is directly proportional to the resistivity, which allows us to infer the sustainability of the quasiparticle picture in these models. 
\begin{table}[h]
\begin{tabular}{|c||c|c|}
\hline
Observable & Lines $\mathcal{N}$, $\overline{\mathcal{N}}$ & Line $\mathcal{P}$ \\ \hline
$c_v$ & $ \sim T + \alpha_1 T^2$ & $ \sim T ^{\frac{1}{4}} + \alpha_2 T^{\frac{1}{2}}$ \\ \hline
$\rho$ & $\sim$ const. + $ \alpha_3 T$ & $\sim$ const. + $\alpha_4 T^{\frac{1}{4}}$\\
\hline
\end{tabular}
\caption{{Specific Heat and Resistivity scaling behaviours (at third order in Kondo coupling) associated with lines of fixed points $\mathcal{N}$, $\overline{\mathcal{N}}$, and $\mathcal{P}$. $\alpha_{1,2,3,4}$ denote proportionality constants.}}
\label{tab_phys_all}
\end{table}

Lines $\mathcal{N}$, $\overline{\mathcal{N}}$ provide the same scaling behaviour as the usual two-channel Kondo model at {two-loop} calculation. This is not altogether surprising, as these lines include the $e_g$ fixed points.
One may imagine these lines as being an extension of the Nozi\`eres like fixed points in the eight-dimensional parameter space. Based on our perturbative analysis, we can thus claim that the low-energy physics described by Lines $\mathcal{N}$, $\overline{\mathcal{N}}$ is precisely the same as the two-channel Kondo model i.e. we obtain the non-Fermi liquid of the two-channel model.

Line $\mathcal{P}$ (and the $p$-only fixed points it includes) is a novel line of fixed points, with distinct non-Fermi liquid behaviour. Specifically, as seen in Table \ref{tab_phys_all}, it has more singular behaviour in its scaling exponents for both the specific heat and resistivity, as compared to the usual two-channel Kondo model. For this line, the scattering rate is much larger ($\sim T^{1/4}$) than the quasiparticle energy ($\sim T$); this is equivalent to stating that the time it takes for the quasiparticle (eigenstate) to decay is much shorter than the quasiparticle (eigenstate's) lifetime. As such, the breakdown of the Landau picture is dramatic along this line, leading to the absence of well-defined quasiparticles and the emergence of an exotic metallic phase with non-trivial low-lying excitations.
We note that the scaling behaviours discussed here are valid up to third order in perturbative renormalization group analysis {and as a consequence, they do not match up with the exponents from more sophisticated (and exact) approaches;
\textcolor{black}{for instance, in the known two-channel Kondo model, Bethe ansatz exact solution yields a logarithmically singular scaling to the specific heat $c_{v} \sim T \ln T$ \cite{andrei_bethe_multi, Hewson_1993}, and boundary CFT \cite{affleck_two_channel_resistivity} predicts the resistivity is of the form $\rho \sim \text{const.} + \sqrt{T}$. }
Higher-order contributions may thus change the exponents reported above, however, we expect them to remain singular (as is the case for the two-channel Kondo model \cite{affleck_cft_review_2}).

\section{Discussions}

In this work, we identified novel non-Fermi liquid states in multipolar quantum impurity systems, where conduction electrons interact with local quadrupolar and octupolar moments. From our perturbative RG analysis, we found that low temperature behaviours of such systems can be governed by more than one possible non-Fermi liquid states depending on the bare parameters of the systems. Such novel behaviors stem from the presence of two kinds of stable fixed points in the RG flow, one corresponding to the two channel Kondo problem and the other a newly discovered non-Fermi liquid state with highly singular physical properties. 
Hence, \textcolor{black}{drawing inspiration from heavy fermion materials where non-thermal parameters can be used to tune the ground state \cite{nat_phys_hfm_review}}, if the bare parameters of the systems are changed by pressure or any other physical means, one {would see a quantum phase transition between the two distinct non-Fermi liquid phases at zero-temperature, and a crossover at low temperatures}.
\textcolor{black}{For instance, applying a hydrostatic pressure \cite{pressure_hfm_super_quad_pr_ti} would be means to compress the FK cage and thus increase the hybridization amplitudes  (while preserving the $T_d$ symmetry of the FK cage).
Another means would be via chemical substitution of the transition metal in (for example) PrTi$_2$Al$_{20}$ with Vanadium; this plays the role of a chemical pressure to compress the unit cell \cite{pr_nfl}.
Both such physical and chemical pressures can thus be used to increase the hybridization amplitudes (and subsequently the Kondo couplings), and allow exploration of the coupling parameter space and the non-Fermi liquid ground states.}

While we solved a single multipolar quantum impurity system, our results imply possibly 
striking consequences for the Kondo lattice model and heavy fermion systems involving ``hidden order''. In particular, the conventional heavy fermion phase diagram must be reimagined as the strong-coupling limit may not merely be Fermi liquid-like, but non-Fermi liquid-like with singular scaling in its response functions. Moreover, this work stresses the importance of incorporating conduction electron orbitals with different symmetries in determining the ultimate low energy ground states. This is highly suggestive that the inclusion of conduction electron orbitals with different symmetries in multipolar quantum impurity systems may be a new avenue to engineer emergent non-Fermi liquid states. 

There are fascinating aspects of this work that still require further investigation. Firstly, the perturbative renormalization group approach evoked here establishes the existence of new non-Fermi liquid states, signified by the non-trivial fixed points in the RG flow. The next natural step would be to confront this problem, in the manner of the conventional Kondo problem, with boundary CFT \cite{affelck_cft_review_impurity, affleck_cft_review_2, affleck_cft_multi, ludwig_exact_multi_review} or Bethe ansatz approaches \cite{bethe_orig_kondo, andrei_bethe_multi, bethe_multi_kondo, Tsvelick_bethe}. 
{It would be interesting to find the exact exponents of the above physical properties' scaling, so as to rigorously identify the nature of the novel non-Fermi liquid.}
Secondly, extension of this single impurity study to a generalized lattice of localized moments would be an interesting direction {to pursue} \cite{Qimiao_kondo_destruction}. 
\textcolor{black}{Finally, it would be intriguing to study the single multipolar impurity with scanning electron microscopy (STM) techniques, which have been historically fruitful endeavours in detecting the Kondo resonance in single-channel Kondo systems \cite{Madhavan567, kondo_resonance_cu_ag}. 
Extension of previous theoretical predictions in multi-channel Kondo systems (which suggest a characteristic multi-channel Kondo resonance \cite{spectral_weight_cox}) to the multipolar impurity model considered in this work, as well as experimental investigations, may help to provide better understanding of the nature of the unusual non-Fermi liquid ground states.}

\section*{Acknowledgements}
This work was supported by NSERC of Canada. Y.B.K. is supported by the Killam Research Fellowship of the Canada Council for the Arts. We thank SungBin Lee for insightful discussions on the molecular orbital bases of the conduction electrons. We are grateful to Wonjune Choi for illuminating discussions on the Kondo problem. We thank Mingxuan Fu for many helpful discussions on the manuscript.

\vfill

\clearpage
\onecolumngrid
\section*{Appendix}
\appendix 
\section{Basis states of multipolar $f$ electrons}
\label{f_gamma_up_basis}

\noindent The ground state of the $T_d$ subjected $f$-electrons is a non-Kramers doublet which, written in the $\ket{J_z}$ basis, is
\begin{equation}
\begin{aligned}
&\Gamma_{3} ^{(1)} = \frac{1}{2} \sqrt{\frac{7}{6}} \ket{4} - \frac{1}{2} \sqrt{\frac{5}{3}} \ket{0} + \frac{1}{2} \sqrt{ \frac{7}{6} } \ket{-4}, ~~~~~~~~~~~~~~~~
&\Gamma_3 ^{(2)} = \frac{1}{\sqrt{2}} \ket{2} + \frac{1}{\sqrt{2}} \ket{-2}.
\end{aligned}
\end{equation}
We can construct a pseudospin-1/2 basis from these $\Gamma_{3g}$ ($ \{ \ket{\uparrow}, \ket{\downarrow} \} $) as
\begin{equation}
\begin{aligned}
&\ket{\uparrow} = \frac{1}{\sqrt{2}} \left[ \ket{\Gamma_{3} ^{(1)}} + i \ket{\Gamma_{3} ^{(2)}} \right], ~~~~~~~~~~~~~~~~
&\ket{\downarrow} = \frac{1}{\sqrt{2}} \left[ i \ket{\Gamma_{3} ^{(1)}} + \ket{\Gamma_{3} ^{(2)}} \right],
\end{aligned}
\end{equation}
which allows the multipolar moments to be neatly written as an effective pseudospin-1/2 operator $\vec{S} = (S^x, S^y, S^z)$ (see main text).

\textcolor{black}
{The 4$f^1$ excited state belongs to the $\Gamma_7$ irrep. with basis states,
\begin{align}
\ket{\Gamma_7 , \pm} = \sqrt{\frac{1}{6}} \ket{ \pm \frac{5}{2} } - \sqrt{\frac{5}{6}} \ket{ \mp \frac{3}{2} }.
\end{align}
}
\section{Slater Koster Tight binding Model} \label{slater_koster}

We model the conduction electrons kinetic term using the Slater-Koster method of tight binding for the $p$ and $e_g$ orbitals located on sublattice A and B of the diamond lattice. 
To simplify the matrix structure of the kinetic term we retain only first nearest neighbour hoppings, and allow only \textit{like-to-like} orbital hopping.
These assumptions still ensure a localized Fermi surface about the zone centre, and yield kinetic eigenvectors (and eigenvalues) that do not mix different orbitals together, as seen below. This ensures that the form of `Kondo' coupling presented in the next section is unaffected by rewriting it in terms of the kinetic eigenvectors.

We present in Fig. \ref{fig_slater_koster_band} a plot of the tight-binding band structure, as well as a plot of a Fermi surface localized about the band centre (the Fermi energy is set at the dotted line in the band structure) showing the localized Fermi surface about the zone centre. Moreover, focussing on the bands closest to the Fermi surface, one can see a set of \textit{nearly} degenerate bands, which have \textit{approximately} the same bandwidth. We note that it is reasonable to take the bands to be degenerate in momentum space, as any non-degeneracy manifests itself in terms of an irrelevant perturbation (in the renormalization group sense) for the two-channel Kondo model \cite{cox_2_channel_stability}.

\begin{figure}[h]
        \includegraphics[width=0.6\textwidth]{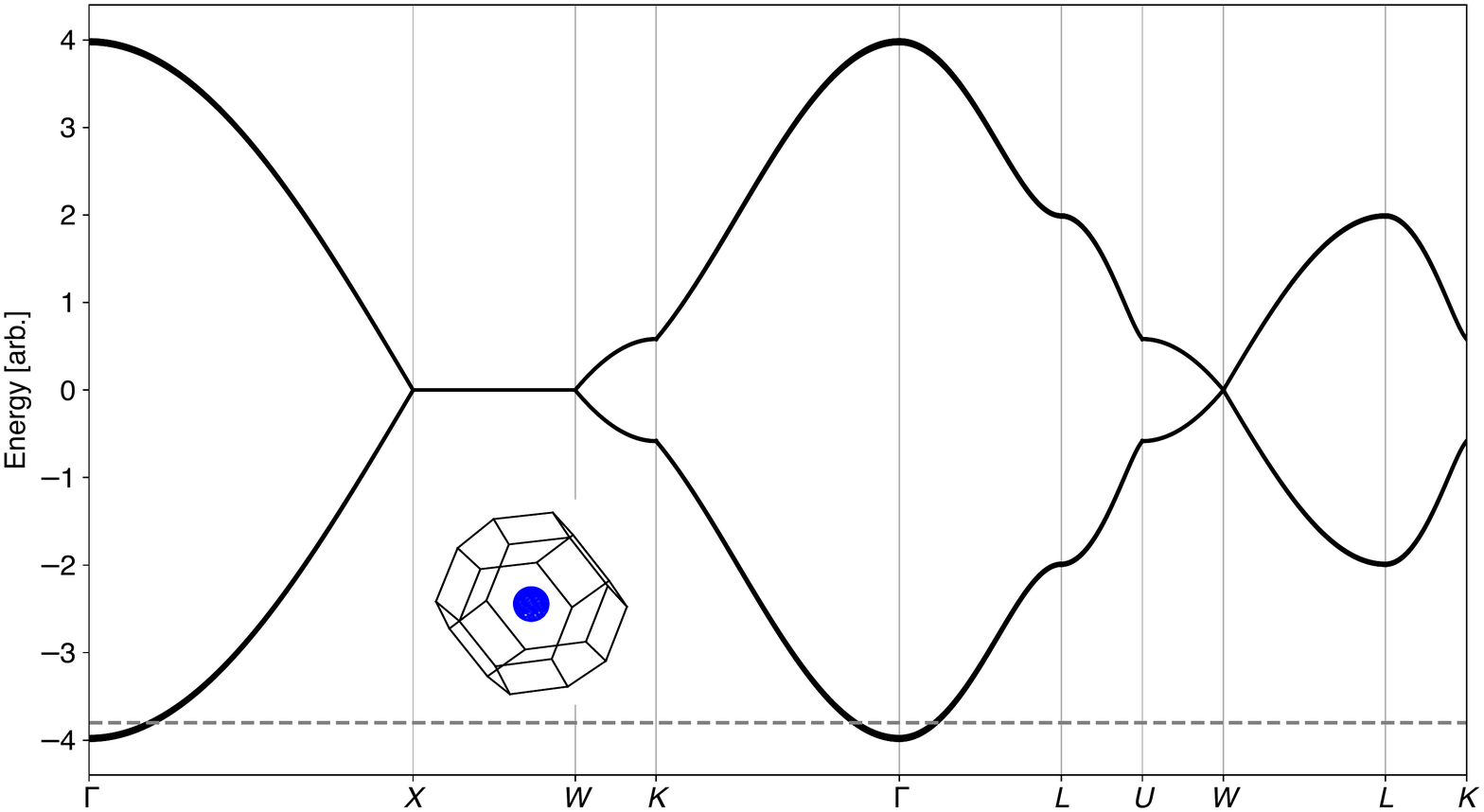} 
          \caption{Slater-Koster Tight binding band structure of $e_g \otimes T_2$ model, and Fermi level indicated by `dotted-line' in band structure for $t=1.5$, $t' = 0$, and $s_{T_2}= p_{T_2} = 1$, $p_e = 1$, $d_e = 0.2$. Inset: Fermi surface (blue) depicted in first Brillouin zone (black).}
          \label{fig_slater_koster_band}
\end{figure}
The corresponding eigenvectors and eigenvalues of the Slater-Koster tight binding model involving $p$ and $e_g$ orbitals, and assuming a Fermi surface localized about the zone centre i.e. $ | \vec{k} | \ll 1$ \\ \\ 
\begin{equation}
\begin{aligned}
& c_{\pm1} = \frac{1}{\sqrt{2}} \left(c_{x,A} \pm c_{x,B} \right), ~~~~~~~~
c_{\pm2} = \frac{1}{\sqrt{2}} \left(c_{y,A} \pm c_{y,B} \right),  ~~~~~~~~
c_{\pm3} = \frac{1}{\sqrt{2}} \left(c_{z,A} \pm c_{z,B} \right), \\
& c_{\pm4} = \frac{1}{\sqrt{2}} \left(c_{x^2 -y^2,A} \pm c_{x^2 -y^2,B} \right), ~~~~~~~~~~~~
c_{\pm5} = \frac{1}{\sqrt{2}} \left(c_{2z^2 - x^2 -y^2,A} \pm c_{2z^2 - x^2 -y^2,B} \right),
\end{aligned}
\label{eq_eigenmodes}
\end{equation}
\begin{align}
& E_{\pm1,\pm2,\pm3} =  \pm s_{T_2} \left(4 - \frac{k^2}{4} \right), ~~~~~~ E_{\pm 4, \pm 5} =  \pm \frac{d_e + 2 p_e}{3} \left(4 - \frac{k^2}{4} \right),
\end{align} \\ \\
where $s_{T_2}$ is the $\sigma$ overlap for the $T_2$ orbitals; and $p_e$ and $d_e$ are the $\pi$ and $\delta$ overlaps for the $e_g$ orbitals. As seen in Eq. \ref{eq_eigenmodes}, the different orbitals do not mix with each other in the tight-binding eigenbasis.
{We stress that the details of the tight-binding model presented here are not employed in this work; the eigenenergies in Fig. \ref{fig_slater_koster_band} merely serve to demonstrate that a localized Fermi pocket is possible with these choice of molecular orbitals. Throughout our calculations, we only utilize the fact that there is a finite density of states at the Fermi level.}
\noindent \section{Symmetry properties of multipolar moments and conduction electrons} \label{app_symm_property}
We consider here the symmetry properties of the multipolar moments, and then discuss conduction electrons. We note the spin of the conduction electrons is denoted by $\vec{\sigma}$. We recall the two generating elements of the $T_d$ group, namely $S_{4z}$ and $C_{31}$. In real $\mathbb{R}^3$ space, the matrix representations of $S_{4z}$ and $C_{31}$ are,\\
\noindent\begin{minipage}{.5\linewidth}
\[
  S_{4z} = 
    \begin{bmatrix}
0 & -1 & 0 \\
1 & 0 & 0 \\
0 & 0 &1\\
  \end{bmatrix}
  \cdot \mathbb{I},
\]
\end{minipage}%
\begin{minipage}{.5\linewidth}
\[
  C_{31} = 
    \begin{bmatrix}
0 & 0 & 1 \\
1 & 0 & 0 \\
0 & 1 & 0 \\
  \end{bmatrix},
\]
\end{minipage} \\ \\
where $\mathbb{I}$ denotes spatial inversion $(x,y,z) \rightarrow (-x, -y, -z)$. We also consider time-reversal ($\Theta$) symmetry. We present in Table \ref{tab_order} how the multipolar moments, and conduction electrons' spin and orbitals transform under the $T_d$ point group and time-reversal ($\Theta$).
\begin{table}[h]
\begin{tabular}{c||c|c|c}
Operator & ${\Theta}$ & $\mathcal{S}_{4z}$ & $\mathcal{C}_{31} $  \\
\hline
$(S^x, S^y, S^z)$ & $\left(S^x, S^y, -S^z \right)$ & $\left(-S^x, S^y, -S^z \right)$ & $\left(-\frac{1}{2}S ^x  + \frac{\sqrt{3}}{2}S ^y, -\frac{\sqrt{3}}{2}S ^x  - \frac{1}{2}S ^y, S^z \right)$ \\ 
\hline
$(\sigma^x, \sigma^y, \sigma^z)$ & $\left(-\sigma^x, -\sigma^y, -\sigma^z \right)$ & $\left(- \sigma^{y}, \sigma^{x}, \sigma^{z} \right)$ & $\left( \sigma^{z}, \sigma^{x}, \sigma^{y} \right)$ \\ 
\hline
$(p_x, p_y, p_z)$ & $\left(p_x, p_y, p_z \right)$ & $\left(p_y, -p_x, -p_z \right)$ & $\left( p_z, p_x, p_y \right)$ \\ 
\hline
$\left(e_{g1}, e_{g2} \right)$ & $\left(e_{g1}, e_{g2} \right)$ & $\left(e_{g1}, - e_{g2} \right)$ & $\left( - \frac{\sqrt{3}}{2} e_{g2} -  \frac{1}{2}  e_{g1}, - \frac{{1}}{2} e_{g2} +  \frac{\sqrt{3}}{2}  e_{g1} \right)$ 
\end{tabular}
\caption{Transformation of multipolar order parameters, and conduction electron degrees of freedom under generating elements of $T_d$ ($\mathcal{S}_{4z}, \ \mathcal{C}_{31}$), and time reversal ($\Theta$). We use the notation of $e_{g1} = (2z^2 - x^2 -y^2)/\sqrt{3}$ and $e_{g2} = x^2 -y^2$ for the $d$-orbitals.}
\label{tab_order}
\end{table}


\section{Pseudospin representation of conduction orbitals} \label{app_pauli_extended}

Here we present the matrices that are used in describing the orbital degrees of freedom in the main text: (i) $\tau$ is the usual Pauli-$2 \times 2$ matrix spanning the vector space $e_g$-orbitals. (ii) $\lambda$ are the usual Gell-Mann $3 \times 3$ matrices spanning the vector space $p$-orbitals. (iii) $\kappa$ are generalized Gell-Mann $5 \times 5$ matrices spanning the vector space of $e_g \otimes p$-orbitals (we only list a subset of the $5 \times 5$ matrices used in this work). The matrices are represented in the following basis where $e_{g1} = (2z^2 - x^2 -y^2)/\sqrt{3}$ and $e_{g2} = x^2 -y^2$. The `...' symbolically represents a complex number.

$\tau \stackrel{.}{=} $ \bordermatrix{~ & e_{g1} & e_{g2} \cr
                  e_{g1} & ... & ...  \cr  
                   e_{g2} & ... & ...  \cr }  \ \ \ \ \ \ \ \ \  $\lambda \stackrel{.}{=} $ \bordermatrix{~ & p_x & p_y & p_z\cr
                  p_x & ... & ... & ... \cr
                  p_y & ... & ... & ... \cr
                  p_z & ... & ... & ... \cr} \ \ \ \ \ \ \ \ \  $\kappa \stackrel{.}{=} $ \bordermatrix{~ & e_{g1} & e_{g2} & p_x & p_y & p_z\cr
                  e_{g1} & ... & ... & ... & ... & ... \cr
                  e_{g2} & ... & ... & ... & ... & ... \cr     
                  p_x & ... & ... & ... & ... & ... \cr
                  p_y & ... & ... & ... & ... & ... \cr
                  p_z & ... & ... & ... & ... & ... \cr}

In this basis, we have,
\begin{equation}  
\tau^{x}\stackrel{.}{=} \left(
\begin{array}{cc}
 0 & 1 \\
 1 & 0 \\
\end{array}
\right); \ \ \ \ 
\tau^{y}\stackrel{.}{=} \left(
\begin{array}{cc}
 0 & -i \\
 i & 0 \\
\end{array}
\right); \ \ \ \ 
\tau^{z}\stackrel{.}{=} \left(
\begin{array}{cc}
 1 & 0 \\
 0 & -1 \\
\end{array}
\right).
\end{equation}  
\noindent\makebox[\linewidth]{\rule{\textwidth}{0.4pt}}
\begin{equation*}
\lambda^{0}\stackrel{.}{=}\left(
\begin{array}{ccc}
 1 & 0 & 0 \\
 0 & 1 & 0 \\
 0 & 0 & 1 \\
\end{array}
\right); \ \ \ \ 
\lambda^{x^2 - y^2}\stackrel{.}{=}\left(
\begin{array}{ccc}
 1 & 0 & 0 \\
 0 & -1 & 0 \\
 0 & 0 & 0 \\
\end{array}
\right);\ \ \ \ 
\lambda^{2z^2 -x^2 - y^2}\stackrel{.}{=} \frac{-1}{\sqrt{3}} \left(
\begin{array}{ccc}
 -1 & 0 & 0 \\
 0 & -1 & 0 \\
 0 & 0 & 2 \\
\end{array}
\right); \\
\end{equation*}
\begin{equation}
\lambda^{xy,r}\stackrel{.}{=}\left(
\begin{array}{ccc}
 0 & 1 & 0 \\
 1 & 0 & 0 \\
 0 & 0 & 0 \\
\end{array}
\right);\ \ \ \ 
\lambda^{yz,r}\stackrel{.}{=}\left(
\begin{array}{ccc}
 0 & 0 & 0 \\
 0 & 0 & 1 \\
 0 & 1 & 0 \\
\end{array}
\right);\ \ \ \ 
\lambda^{xz,r}\stackrel{.}{=}\left(
\begin{array}{ccc}
 0 & 0 & 1 \\
 0 & 0 & 0 \\
 1 & 0 & 0 \\
\end{array}
\right);
\end{equation}
\begin{equation*}
\lambda^{xy,i}\stackrel{.}{=}\left(
\begin{array}{ccc}
 0 & -i & 0 \\
 i & 0 & 0 \\
 0 & 0 & 0 \\
\end{array}
\right);\ \ \ \ 
\lambda^{yz,i}\stackrel{.}{=}\left(
\begin{array}{ccc}
 0 & 0 & 0 \\
 0 & 0 & -i \\
 0 & i & 0 \\
\end{array}
\right);\ \ \ \ 
\lambda^{xz,i}\stackrel{.}{=}\left(
\begin{array}{ccc}
 0 & 0 & -i \\
 0 & 0 & 0 \\
 i & 0 & 0 \\
\end{array}
\right).
\end{equation*}
\noindent\makebox[\linewidth]{\rule{\textwidth}{0.4pt}}
\begin{equation*}
\kappa^{1x,r} \stackrel{.}{=} \left(
\begin{array}{ccccc}
 0 & 0 & 1 & 0 & 0 \\
 0 & 0 & 0 & 0 & 0 \\
 1 & 0 & 0 & 0 & 0 \\
 0 & 0 & 0 & 0 & 0 \\
 0 & 0 & 0 & 0 & 0 \\
\end{array}
\right);\ \ \ \ 
\kappa^{1y,r} \stackrel{.}{=}  \left(
\begin{array}{ccccc}
 0 & 0 & 0 & 1 & 0 \\
 0 & 0 & 0 & 0 & 0 \\
 0 & 0 & 0 & 0 & 0 \\
 1 & 0 & 0 & 0 & 0 \\
 0 & 0 & 0 & 0 & 0 \\
\end{array}
\right);\ \ \ \ 
\kappa^{1z,r} \stackrel{.}{=} \left(
\begin{array}{ccccc}
 0 & 0 & 0 & 0 & 1 \\
 0 & 0 & 0 & 0 & 0 \\
 0 & 0 & 0 & 0 & 0 \\
 0 & 0 & 0 & 0 & 0 \\
 1 & 0 & 0 & 0 & 0 \\
\end{array}
\right)
\kappa^{2x,r} \stackrel{.}{=} \left(
\begin{array}{ccccc}
 0 & 0 & 0 & 0 & 0 \\
 0 & 0 & 1 & 0 & 0 \\
 0 & 1 & 0 & 0 & 0 \\
 0 & 0 & 0 & 0 & 0 \\
 0 & 0 & 0 & 0 & 0 \\
\end{array}
\right);\ \ \ \ 
\end{equation*}
\begin{equation}
\kappa^{2y,r} \stackrel{.}{=}  \left(
\begin{array}{ccccc}
 0 & 0 & 0 & 0 & 0 \\
 0 & 0 & 0 & 1 & 0 \\
 0 & 0 & 0 & 0 & 0 \\
 0 & 1 & 0 & 0 & 0 \\
 0 & 0 & 0 & 0 & 0 \\
\end{array}
\right);\ \ \ \ 
\kappa^{2z,r} \stackrel{.}{=} \left(
\begin{array}{ccccc}
 0 & 0 & 0 & 0 & 0 \\
 0 & 0 & 0 & 0 & 1 \\
 0 & 0 & 0 & 0 & 0 \\
 0 & 0 & 0 & 0 & 0 \\
 0 & 1 & 0 & 0 & 0 \\
\end{array}
\right); \ \ \ 
\kappa^{1x,i} \stackrel{.}{=} \left(
\begin{array}{ccccc}
 0 & 0 & -i & 0 & 0 \\
 0 & 0 & 0 & 0 & 0 \\
 i & 0 & 0 & 0 & 0 \\
 0 & 0 & 0 & 0 & 0 \\
 0 & 0 & 0 & 0 & 0 \\
\end{array}
\right);\ \ \ 
\kappa^{1y,i} \stackrel{.}{=}  \left(
\begin{array}{ccccc}
 0 & 0 & 0 & -i & 0 \\
 0 & 0 & 0 & 0 & 0 \\
 0 & 0 & 0 & 0 & 0 \\
 i & 0 & 0 & 0 & 0 \\
 0 & 0 & 0 & 0 & 0 \\
\end{array}
\right);\ \ \
\end{equation}
\begin{equation*}
\kappa^{1z,i} \stackrel{.}{=} \left(
\begin{array}{ccccc}
 0 & 0 & 0 & 0 & -i \\
 0 & 0 & 0 & 0 & 0 \\
 0 & 0 & 0 & 0 & 0 \\
 0 & 0 & 0 & 0 & 0 \\
 i & 0 & 0 & 0 & 0 \\
\end{array}
\right);\ \ \
\kappa^{2x,i} \stackrel{.}{=} \left(
\begin{array}{ccccc}
 0 & 0 & 0 & 0 & 0 \\
 0 & 0 & -i & 0 & 0 \\
 0 & i & 0 & 0 & 0 \\
 0 & 0 & 0 & 0 & 0 \\
 0 & 0 & 0 & 0 & 0 \\
\end{array}
\right);\ \ \
\kappa^{2y,i} \stackrel{.}{=}  \left(
\begin{array}{ccccc}
 0 & 0 & 0 & 0 & 0 \\
 0 & 0 & 0 & -i & 0 \\
 0 & 0 & 0 & 0 & 0 \\
 0 & i & 0 & 0 & 0 \\
 0 & 0 & 0 & 0 & 0 \\
\end{array}
\right);\ \ \
\kappa^{2z,i} \stackrel{.}{=} \left(
\begin{array}{ccccc}
 0 & 0 & 0 & 0 & 0 \\
 0 & 0 & 0 & 0 & -i \\
 0 & 0 & 0 & 0 & 0 \\
 0 & 0 & 0 & 0 & 0 \\
 0 & i & 0 & 0 & 0 \\
\end{array}
\right).
\end{equation*}

\section{Microscopic Inter-site Anderson model} 
\label{si_anderson_model}
\textcolor{black}{
As discussed in the main text, we consider a four Al-atom tetrahedron cage which has the same $T_d$ point group.
For the sake of clarity, we consider a Pr atom centred at the origin (0,0,0) and with the four Al atoms on the corners of the cube (to form the tetrahedron): $(1): (1,1,1)$, $(2): (1,-1,-1)$, $(3): (-1,1,-1)$, and $(4): (-1,-1,1)$.
Under the applied transformation using the symmetry generators of the $T_d$ group, the four-sites get mapped to each other.
Under $\mathcal{C}_{31}:$ $(1) \rightarrow (1)$, $(2) \rightarrow (4)$, $(3) \rightarrow (2)$, and $(4) \rightarrow (3)$.
Under $\mathcal{S}_{4z}$: $(1) \rightarrow (3)$, $(2) \rightarrow (1)$, $(3) \rightarrow (4)$, and $(2) \rightarrow (3)$.
The following Table \ref{si_tab_hybr_elements} lists the ninety-six hybridization matrix elements. As seen, they are composed of four independent parameters: $\mathcal{V}_{1,2,3,4}$.
}
\renewcommand{\arraystretch}{2.0}
\begin{longtable}{c|c|c|c|c|c}
\caption{Inter-site hybridization matrix elements, $V_{\Delta \omega j \mathcal{P} \sigma}$.}  \\
\hline
$f^2$ State: $\Delta$ & $f^1$ State:  $\omega$ & Site $j$ & Atomic orbital: $\mathcal{P}$ & c-spin: $\sigma$ & $V_{\Delta \omega j \mathcal{P} \sigma}$ \\
\hline
 (1) & + & 1 & $x$ & $\uparrow$ & $ \frac{1}{4} \left(i \sqrt{3}+1\right) (\mathcal{V}_2-i \mathcal{V}_1+i \mathcal{V}_4+\mathcal{V}_3) $ \\
 (1) & + & 1 & $x$ & $\downarrow$ & $ \frac{1}{4} \left(-i+\sqrt{3}\right) (-i \mathcal{V}_2+\mathcal{V}_1+\mathcal{V}_4+i \mathcal{V}_3) $ \\
 (1) & + & 1 & $y$ & $\uparrow$ & $ \frac{1}{4} \left(i+\sqrt{3}\right) (\mathcal{V}_2+\mathcal{V}_1+\mathcal{V}_4+\mathcal{V}_3) $ \\
 (1) & + & 1 & $y$ & $\downarrow$ & $ \frac{1}{4} i \left(i+\sqrt{3}\right) (\mathcal{V}_2-\mathcal{V}_1+\mathcal{V}_4-\mathcal{V}_3) $ \\
 (1) & + & 1 & $z$ & $\uparrow$ & $ \mathcal{V}_1 $ \\
 (1) & + & 1 & $z$ & $\downarrow$ & $ \mathcal{V}_2 $ \\
 (1) & + & 2 & $x$ & $\uparrow$ & $ \frac{1}{4} \left(i \sqrt{3}+1\right) (\mathcal{V}_2+i \mathcal{V}_1-i \mathcal{V}_4+\mathcal{V}_3) $ \\
 (1) & + & 2 & $x$ & $\downarrow$ & $ -\frac{1}{4} i \left(-i+\sqrt{3}\right) (\mathcal{V}_2-i (\mathcal{V}_1+\mathcal{V}_4-i \mathcal{V}_3)) $ \\
 (1) & + & 2 & $y$ & $\uparrow$ & $ \frac{1}{4} \left(i+\sqrt{3}\right) (\mathcal{V}_2-\mathcal{V}_1-\mathcal{V}_4+\mathcal{V}_3) $ \\
 (1) & + & 2 & $y$ & $\downarrow$ & $ \frac{1}{4} \left(1-i \sqrt{3}\right) (\mathcal{V}_2+\mathcal{V}_1-\mathcal{V}_4-\mathcal{V}_3) $ \\
 (1) & + & 2 & $z$ & $\uparrow$ & $ \mathcal{V}_4 $ \\
 (1) & + & 2 & $z$ & $\downarrow$ & $ \mathcal{V}_3 $ \\
 (1) & + & 3 & $x$ & $\uparrow$ & $ \frac{1}{4} \left(i \sqrt{3}+1\right) (\mathcal{V}_2+i \mathcal{V}_1-i \mathcal{V}_4+\mathcal{V}_3) $ \\
 (1) & + & 3 & $x$ & $\downarrow$ & $ \frac{1}{4} \left(-i+\sqrt{3}\right) (i \mathcal{V}_2+\mathcal{V}_1+\mathcal{V}_4-i \mathcal{V}_3) $ \\
 (1) & + & 3 & $y$ & $\uparrow$ & $ \frac{1}{4} \left(i+\sqrt{3}\right) (\mathcal{V}_2-\mathcal{V}_1-\mathcal{V}_4+\mathcal{V}_3) $ \\
 (1) & + & 3 & $y$ & $\downarrow$ & $ \frac{1}{4} i \left(i+\sqrt{3}\right) (\mathcal{V}_2+\mathcal{V}_1-\mathcal{V}_4-\mathcal{V}_3) $ \\
 (1) & + & 3 & $z$ & $\uparrow$ & $ -\mathcal{V}_4 $ \\
 (1) & + & 3 & $z$ & $\downarrow$ & $ \mathcal{V}_3 $ \\
 (1) & + & 4 & $x$ & $\uparrow$ & $ \frac{1}{4} \left(i \sqrt{3}+1\right) (\mathcal{V}_2-i \mathcal{V}_1+i \mathcal{V}_4+\mathcal{V}_3) $ \\
 (1) & + & 4 & $x$ & $\downarrow$ & $ \frac{1}{4} \left(i \sqrt{3}+1\right) (\mathcal{V}_2+i (\mathcal{V}_1+\mathcal{V}_4+i \mathcal{V}_3)) $ \\
 (1) & + & 4 & $y$ & $\uparrow$ & $ \frac{1}{4} \left(i+\sqrt{3}\right) (\mathcal{V}_2+\mathcal{V}_1+\mathcal{V}_4+\mathcal{V}_3) $ \\
 (1) & + & 4 & $y$ & $\downarrow$ & $ \frac{1}{4} i \left(i+\sqrt{3}\right) (-\mathcal{V}_2+\mathcal{V}_1-\mathcal{V}_4+\mathcal{V}_3) $ \\
 (1) & + & 4 & $z$ & $\uparrow$ & $ -\mathcal{V}_1 $ \\
 (1) & + & 4 & $z$ & $\downarrow$ & $ \mathcal{V}_2 $ \\
 (1) & - & 1 & $x$ & $\uparrow$ & $ \frac{1}{4} \left(i \sqrt{3}+1\right) (\mathcal{V}_2-i (\mathcal{V}_1+\mathcal{V}_4-i \mathcal{V}_3)) $ \\
 (1) & - & 1 & $x$ & $\downarrow$ & $ \frac{1}{4} \left(-i+\sqrt{3}\right) (-i \mathcal{V}_2+\mathcal{V}_1-\mathcal{V}_4-i \mathcal{V}_3) $ \\
 (1) & - & 1 & $y$ & $\uparrow$ & $ \frac{1}{4} \left(1-i \sqrt{3}\right) (\mathcal{V}_2+\mathcal{V}_1-\mathcal{V}_4-\mathcal{V}_3) $ \\
 (1) & - & 1 & $y$ & $\downarrow$ & $ \frac{1}{4} \left(i+\sqrt{3}\right) (\mathcal{V}_2-\mathcal{V}_1-\mathcal{V}_4+\mathcal{V}_3) $ \\
 (1) & - & 1 & $z$ & $\uparrow$ & $ \mathcal{V}_3 $ \\
 (1) & - & 1 & $z$ & $\downarrow$ & $ \mathcal{V}_4 $ \\
 (1) & - & 2 & $x$ & $\uparrow$ & $ -\frac{1}{4} \left(-i+\sqrt{3}\right) (-i \mathcal{V}_2+\mathcal{V}_1+\mathcal{V}_4+i \mathcal{V}_3) $ \\
 (1) & - & 2 & $x$ & $\downarrow$ & $ \frac{1}{4} \left(-i+\sqrt{3}\right) (-i \mathcal{V}_2-\mathcal{V}_1+\mathcal{V}_4-i \mathcal{V}_3) $ \\
 (1) & - & 2 & $y$ & $\uparrow$ & $ \frac{1}{4} i \left(i+\sqrt{3}\right) (\mathcal{V}_2-\mathcal{V}_1+\mathcal{V}_4-\mathcal{V}_3) $ \\
 (1) & - & 2 & $y$ & $\downarrow$ & $ \frac{1}{4} \left(i+\sqrt{3}\right) (\mathcal{V}_2+\mathcal{V}_1+\mathcal{V}_4+\mathcal{V}_3) $ \\
 (1) & - & 2 & $z$ & $\uparrow$ & $ \mathcal{V}_2 $ \\
 (1) & - & 2 & $z$ & $\downarrow$ & $ \mathcal{V}_1 $ \\
 (1) & - & 3 & $x$ & $\uparrow$ & $ \frac{1}{4} \left(-i+\sqrt{3}\right) (-i \mathcal{V}_2+\mathcal{V}_1+\mathcal{V}_4+i \mathcal{V}_3) $ \\
 (1) & - & 3 & $x$ & $\downarrow$ & $ \frac{1}{4} \left(-i+\sqrt{3}\right) (-i \mathcal{V}_2-\mathcal{V}_1+\mathcal{V}_4-i \mathcal{V}_3) $ \\
 (1) & - & 3 & $y$ & $\uparrow$ & $ \frac{1}{4} \left(1-i \sqrt{3}\right) (\mathcal{V}_2-\mathcal{V}_1+\mathcal{V}_4-\mathcal{V}_3) $ \\
 (1) & - & 3 & $y$ & $\downarrow$ & $ \frac{1}{4} \left(i+\sqrt{3}\right) (\mathcal{V}_2+\mathcal{V}_1+\mathcal{V}_4+\mathcal{V}_3) $ \\
 (1) & - & 3 & $z$ & $\uparrow$ & $ \mathcal{V}_2 $ \\
 (1) & - & 3 & $z$ & $\downarrow$ & $ -\mathcal{V}_1 $ \\
 (1) & - & 4 & $x$ & $\uparrow$ & $ -\frac{1}{4} i \left(-i+\sqrt{3}\right) (\mathcal{V}_2-i (\mathcal{V}_1+\mathcal{V}_4-i \mathcal{V}_3)) $ \\
 (1) & - & 4 & $x$ & $\downarrow$ & $ \frac{1}{4} \left(-i+\sqrt{3}\right) (-i \mathcal{V}_2+\mathcal{V}_1-\mathcal{V}_4-i \mathcal{V}_3) $ \\
 (1) & - & 4 & $y$ & $\uparrow$ & $ \frac{1}{4} i \left(i+\sqrt{3}\right) (\mathcal{V}_2+\mathcal{V}_1-\mathcal{V}_4-\mathcal{V}_3) $ \\
 (1) & - & 4 & $y$ & $\downarrow$ & $ \frac{1}{4} \left(i+\sqrt{3}\right) (\mathcal{V}_2-\mathcal{V}_1-\mathcal{V}_4+\mathcal{V}_3) $ \\
 (1) & - & 4 & $z$ & $\uparrow$ & $ \mathcal{V}_3 $ \\
 (1) & - & 4 & $z$ & $\downarrow$ & $ -\mathcal{V}_4 $ \\
 (2) & + & 1 & $x$ & $\uparrow$ & $ \frac{1}{4} \left(i+\sqrt{3}\right) (\mathcal{V}_2-\mathcal{V}_1-\mathcal{V}_4+\mathcal{V}_3) $ \\
 (2) & + & 1 & $x$ & $\downarrow$ & $ \frac{1}{4} \left(i+\sqrt{3}\right) (\mathcal{V}_2+\mathcal{V}_1-\mathcal{V}_4-\mathcal{V}_3) $ \\
 (2) & + & 1 & $y$ & $\uparrow$ & $ -\frac{1}{4} \left(-i+\sqrt{3}\right) (i \mathcal{V}_2-\mathcal{V}_1+\mathcal{V}_4+i \mathcal{V}_3) $ \\
 (2) & + & 1 & $y$ & $\downarrow$ & $ -\frac{1}{4} \left(-i+\sqrt{3}\right) (\mathcal{V}_2-i (\mathcal{V}_1+\mathcal{V}_4-i \mathcal{V}_3)) $ \\
 (2) & + & 1 & $z$ & $\uparrow$ & $ \mathcal{V}_4 $ \\
 (2) & + & 1 & $z$ & $\downarrow$ & $ i \mathcal{V}_3 $ \\
 (2) & + & 2 & $x$ & $\uparrow$ & $ \frac{1}{4} \left(i+\sqrt{3}\right) (\mathcal{V}_2+\mathcal{V}_1+\mathcal{V}_4+\mathcal{V}_3) $ \\
 (2) & + & 2 & $x$ & $\downarrow$ & $ \frac{1}{4} \left(i+\sqrt{3}\right) (\mathcal{V}_2-\mathcal{V}_1+\mathcal{V}_4-\mathcal{V}_3) $ \\
 (2) & + & 2 & $y$ & $\uparrow$ & $ -\frac{1}{4} \left(i \sqrt{3}+1\right) (\mathcal{V}_2-i \mathcal{V}_1+i \mathcal{V}_4+\mathcal{V}_3) $ \\
 (2) & + & 2 & $y$ & $\downarrow$ & $ \frac{1}{4} \left(-i+\sqrt{3}\right) (\mathcal{V}_2+i (\mathcal{V}_1+\mathcal{V}_4+i \mathcal{V}_3)) $ \\
 (2) & + & 2 & $z$ & $\uparrow$ & $ -\mathcal{V}_1 $ \\
 (2) & + & 2 & $z$ & $\downarrow$ & $ i \mathcal{V}_2 $ \\
 (2) & + & 3 & $x$ & $\uparrow$ & $ \frac{1}{4} \left(i+\sqrt{3}\right) (\mathcal{V}_2+\mathcal{V}_1+\mathcal{V}_4+\mathcal{V}_3) $ \\
 (2) & + & 3 & $x$ & $\downarrow$ & $ \frac{1}{4} \left(i+\sqrt{3}\right) (-\mathcal{V}_2+\mathcal{V}_1-\mathcal{V}_4+\mathcal{V}_3) $ \\
 (2) & + & 3 & $y$ & $\uparrow$ & $ -\frac{1}{4} \left(-i+\sqrt{3}\right) (i \mathcal{V}_2+\mathcal{V}_1-\mathcal{V}_4+i \mathcal{V}_3) $ \\
 (2) & + & 3 & $y$ & $\downarrow$ & $ -\frac{1}{4} \left(-i+\sqrt{3}\right) (\mathcal{V}_2+i (\mathcal{V}_1+\mathcal{V}_4+i \mathcal{V}_3)) $ \\
 (2) & + & 3 & $z$ & $\uparrow$ & $ \mathcal{V}_1 $ \\
 (2) & + & 3 & $z$ & $\downarrow$ & $ i \mathcal{V}_2 $ \\
 (2) & + & 4 & $x$ & $\uparrow$ & $ \frac{1}{4} \left(i+\sqrt{3}\right) (\mathcal{V}_2-\mathcal{V}_1-\mathcal{V}_4+\mathcal{V}_3) $ \\
 (2) & + & 4 & $x$ & $\downarrow$ & $ -\frac{1}{4} \left(i+\sqrt{3}\right) (\mathcal{V}_2+\mathcal{V}_1-\mathcal{V}_4-\mathcal{V}_3) $ \\
 (2) & + & 4 & $y$ & $\uparrow$ & $ -\frac{1}{4} \left(-i+\sqrt{3}\right) (i \mathcal{V}_2-\mathcal{V}_1+\mathcal{V}_4+i \mathcal{V}_3) $ \\
 (2) & + & 4 & $y$ & $\downarrow$ & $ \frac{1}{4} \left(-i+\sqrt{3}\right) (\mathcal{V}_2-i (\mathcal{V}_1+\mathcal{V}_4-i \mathcal{V}_3)) $ \\
 (2) & + & 4 & $z$ & $\uparrow$ & $ -\mathcal{V}_4 $ \\
 (2) & + & 4 & $z$ & $\downarrow$ & $ i \mathcal{V}_3 $ \\
 (2) & - & 1 & $x$ & $\uparrow$ & $ -\frac{1}{4} \left(i+\sqrt{3}\right) (\mathcal{V}_2-\mathcal{V}_1+\mathcal{V}_4-\mathcal{V}_3) $ \\
 (2) & - & 1 & $x$ & $\downarrow$ & $ -\frac{1}{4} \left(i+\sqrt{3}\right) (\mathcal{V}_2+\mathcal{V}_1+\mathcal{V}_4+\mathcal{V}_3) $ \\
 (2) & - & 1 & $y$ & $\uparrow$ & $ \frac{1}{4} \left(-i+\sqrt{3}\right) (\mathcal{V}_2+i (\mathcal{V}_1+\mathcal{V}_4+i \mathcal{V}_3)) $ \\
 (2) & - & 1 & $y$ & $\downarrow$ & $ -\frac{1}{4} \left(-i+\sqrt{3}\right) (i \mathcal{V}_2+\mathcal{V}_1-\mathcal{V}_4+i \mathcal{V}_3) $ \\
 (2) & - & 1 & $z$ & $\uparrow$ & $ i \mathcal{V}_2 $ \\
 (2) & - & 1 & $z$ & $\downarrow$ & $ -\mathcal{V}_1 $ \\
 (2) & - & 2 & $x$ & $\uparrow$ & $ -\frac{1}{4} \left(i+\sqrt{3}\right) (\mathcal{V}_2+\mathcal{V}_1-\mathcal{V}_4-\mathcal{V}_3) $ \\
 (2) & - & 2 & $x$ & $\downarrow$ & $ -\frac{1}{4} \left(i+\sqrt{3}\right) (\mathcal{V}_2-\mathcal{V}_1-\mathcal{V}_4+\mathcal{V}_3) $ \\
 (2) & - & 2 & $y$ & $\uparrow$ & $ -\frac{1}{4} \left(-i+\sqrt{3}\right) (\mathcal{V}_2-i (\mathcal{V}_1+\mathcal{V}_4-i \mathcal{V}_3)) $ \\
 (2) & - & 2 & $y$ & $\downarrow$ & $ \frac{1}{4} \left(-i+\sqrt{3}\right) (-i \mathcal{V}_2+\mathcal{V}_1-\mathcal{V}_4-i \mathcal{V}_3) $ \\
 (2) & - & 2 & $z$ & $\uparrow$ & $ i \mathcal{V}_3 $ \\
 (2) & - & 2 & $z$ & $\downarrow$ & $ \mathcal{V}_4 $ \\
 (2) & - & 3 & $x$ & $\uparrow$ & $ \frac{1}{4} \left(i+\sqrt{3}\right) (\mathcal{V}_2+\mathcal{V}_1-\mathcal{V}_4-\mathcal{V}_3) $ \\
 (2) & - & 3 & $x$ & $\downarrow$ & $ \frac{1}{4} \left(i+\sqrt{3}\right) (-\mathcal{V}_2+\mathcal{V}_1+\mathcal{V}_4-\mathcal{V}_3) $ \\
 (2) & - & 3 & $y$ & $\uparrow$ & $ \frac{1}{4} \left(-i+\sqrt{3}\right) (\mathcal{V}_2-i (\mathcal{V}_1+\mathcal{V}_4-i \mathcal{V}_3)) $ \\
 (2) & - & 3 & $y$ & $\downarrow$ & $ \frac{1}{4} \left(-i+\sqrt{3}\right) (-i \mathcal{V}_2+\mathcal{V}_1-\mathcal{V}_4-i \mathcal{V}_3) $ \\
 (2) & - & 3 & $z$ & $\uparrow$ & $ i \mathcal{V}_3 $ \\
 (2) & - & 3 & $z$ & $\downarrow$ & $ -\mathcal{V}_4 $ \\
 (2) & - & 4 & $x$ & $\uparrow$ & $ \frac{1}{4} \left(i+\sqrt{3}\right) (\mathcal{V}_2-\mathcal{V}_1+\mathcal{V}_4-\mathcal{V}_3) $ \\
 (2) & - & 4 & $x$ & $\downarrow$ & $ -\frac{1}{4} \left(i+\sqrt{3}\right) (\mathcal{V}_2+\mathcal{V}_1+\mathcal{V}_4+\mathcal{V}_3) $ \\
 (2) & - & 4 & $y$ & $\uparrow$ & $ -\frac{1}{4} \left(-i+\sqrt{3}\right) (\mathcal{V}_2+i (\mathcal{V}_1+\mathcal{V}_4+i \mathcal{V}_3)) $ \\
 (2) & - & 4 & $y$ & $\downarrow$ & $ \frac{1}{4} \left(-i+\sqrt{3}\right) (-i \mathcal{V}_2-\mathcal{V}_1+\mathcal{V}_4-i \mathcal{V}_3) $ \\
 (2) & - & 4 & $z$ & $\uparrow$ & $ i \mathcal{V}_2 $ \\
 (2) & - & 4 & $z$ & $\downarrow$ & $ \mathcal{V}_1 $ 
 \label{si_tab_hybr_elements}
  \end{longtable}

\section{Schrieffer-Wolff (SW) transformation}
\label{si_sw_transform}
\textcolor{black}{
The SW transformation is a means to perturbatively diagonalize a general interacting Hamiltonian using a unitary transformation.
The transformation also results in renormalizing the parameters of the Hamiltonian, and as such can be thought of as a renormalization procedure.
Since this a standard second-order canonical transformation \cite{Phillips, haq2019explicit}, we neglect the step-by-step procedure, and rather choose to highlight the key steps and results that make it different from its usual implementation in the original Anderson model.
}

The Hamiltonian is split into two pieces: (i) $H_0$ which contains the terms acting independently on the low-energy ($f^2$) and high-energy ($f^1$) subspaces, and (ii) $H_1$ which connects the low-energy and high-energy subspaces i.e. $H = H_0 + H_1$, where
\begin{align}
H_0 = \sum_{\alpha, \sigma} \epsilon_{\alpha} c_{\alpha \sigma} ^{\dag} c_{\alpha \sigma} + \epsilon_f \sum_{\Delta} \ket{f^2, \Delta} \bra{f^2, \Delta},
\end{align}
where $\alpha$ sums over the molecular orbitals, $\sigma$ is the spin of the conduction electrons, $\Delta$ sums over the two $f^2$ non-Kramers ground states, and $\epsilon_f$ is the energy level spacing between the $f^2$ ground state and the excited $f^1$ state. The $f^1$ state is placed at zero-energy so as to ignore its presence in $H_0$. 
The use of molecular orbitals, instead of the atomic orbitals on each site, is beneficial as it renders the conduction electron kinetic term diagonal; 
the tight-binding hopping matrix on the single tetrahedron has molecular orbitals (eigenstates) decoupled as: $A_1 \oplus E \oplus T_1 \oplus 2T_2$, where we use the $T_d$ irrep notation. Here $E$ and $T_2$ have molecular $e_g$ and $p$ as its basis functions. 

The connecting-to the high energy space term is
\begin{align}
H_1 = \sum_{\Delta, \omega, \alpha, \sigma} \tilde{V}_{\Delta \omega \alpha \sigma} \ket{f^2, \Delta} \bra{f^1, \omega} c_{\alpha \sigma} + {\rm{h.c.}}
\end{align}
where $\tilde{V}_{\Delta \omega \alpha \sigma}$ involves linear combinations of the atomic hybridization matrix elements presented in Table \ref{si_tab_hybr_elements}, $\omega$ sums over the $f^1$ states, and we again employ the conduction electron molecular orbitals, $\alpha$.
Due to the Hubbard operators not obeying standard fermion/boson commutation rules, it is cumbersome to employ them directly in the SW transformation.
Instead, we employ the pseudo-particle method \cite{cox_zawadowski_book},
\begin{align}
b_{\omega} ^{\dag} \ket{0} = \ket{f^1, \omega}, ~~~~~~~~~~~~~ f_{\Delta} ^{\dag} \ket{0} = \ket{f^2, \Delta}
\end{align}
where $\ket{0}$ is the vacuum, $b_{\omega}^{\dag}$ is a pseudo-boson operator that creates the $\ket{f^1, \omega} $ state, and $f_{\Delta} ^{\dag}$ is a pseudo-fermion operator that creates the $\ket{f^2, \Delta} $ state.
Indeed one can interchange the usage of a pseudo-boson and pseudo-fermion for the two configurations, as we must only ensure that one configuration is fermionic while the other is bosonic, so as to retain the overall fermionic structure of hybridization. We can then rewrite the hybridization term as,
\begin{align}
H_1 = \sum_{\Delta, \omega, \alpha, \sigma} \tilde{V}_{\Delta \omega \alpha \sigma} f_{\Delta}^{\dag} b_{\omega} c_{\alpha \sigma} + {\rm{h.c.}}
\end{align}
We now apply the SW transformation i.e. $H_{\rm eff} = {\rm e}^{S} (H_0 + H_1) {\rm e}^{-S} = H_0 + \frac{1}{2} \left[ S, H_1 \right]$.
We find that the generator of the transformation, S, is
\begin{align}
S = \sum_{\Delta, \omega, \alpha, \sigma} \frac{\tilde{V}_{\Delta \omega \alpha \sigma}}{\epsilon_f - \epsilon_{\alpha} } f_{\Delta}^{\dag} b_{\omega} c_{\alpha \sigma} - {\rm{h.c.}}
\end{align}
Working through the commutators, we arrive at the effective low-energy Hamiltonian,
\begin{align}
H_{\rm eff} = - \frac{1}{2} \sum_{\Delta, \Delta ', \omega, \alpha, \alpha ', \sigma, \sigma ' }   \frac{\tilde{V}_{\Delta \omega \alpha \sigma} \tilde{V}^*_{\Delta ' \omega \alpha ' \sigma '} }{\epsilon_f - \epsilon_{\alpha} } f_{\Delta}^{\dag} f_{\Delta '} c_{\alpha ' \sigma '} ^{\dag} c_{\alpha \sigma} +  {\rm{h.c.}}
\label{eq_si_sw_kondo}
\end{align}
For clarity, we note that the SW transformation generates terms of the form $~ b_{\omega} ^{\dag} b_{\omega '} F \left(f, f^{\dag}, c, c^{\dag} \right)$, where $F(...)$ is a quartic function in the $f$ and $c$ operators; however, in the low-energy subspace, the $f^1$ state is unoccupied and so these terms vanish in the low-energy theory.
Equation \ref{eq_si_sw_kondo} leads to the `Kondo' Hamiltonians presented in the main text, which were based on symmetry analysis.

\section{Perturbative renormalization group approach for multipolar quantum impurity models}  \label{sec_method_rg}

In this section, we discuss the general strategy used to tackle multipolar quantum impurity systems and obtain the renormalization group (RG) $\beta$-function for the various coupling constants. Specifically, we are interested in computing the scattering amplitude which is obtained by perturbatively computing the four-point correlation function, $\Gamma$.
\\

We consider a general Hamiltonian of the system, without specifying the particular orbitals, as \\
${H} = \sum_{k,a,\alpha} \epsilon_k c^{\dag}_{k,a,\alpha} c_{k,a,\alpha} + H_{\text{Kondo}}[c^{\dag}, c, \vec{S}]$, 
where $a$ and $\alpha$ refer to the orbital and spin index of the conduction electron. $H_{\text{Kondo}}[c^{\dag}, c, \vec{S}]$ denotes the `Kondo' interaction and is quartic in fermionic operators; the specific form depends on the orbitals of interest. We take the conduction electron bands to be degenerate, and we make the further simplification (in the spirit of Anderson \cite{Anderson_poor_1970}) of a constant density of states: $\rho(E) = \rho$, where $-D < E < D$ is the energy window and $D$ is the bandwidth. We note that choosing different density of states for the different orbitals does not modify the subsequent RG flow equations. The Fermi level is taken as the zero-energy level. We also represent the pseudospin-1/2 operator in terms of Abrikosov pseudofermions,
$\vec{S} = \frac{1}{2} \sum_{\mu = \pm} f^{\dag}_{\mu} \vec{\sigma}_{\mu, \nu} f_{\nu}$,
where $f$ denote the pseudofermions. As is typical when using Abrikosov pseudofermions, the Hilbert space gets enlarged by their introduction through inclusion of unphysical doubly occupied and unoccupied states. To correct for this, we employ the standard Popov-Fedotov trick of introducing a complex chemical potential ($\lambda = \frac{i \pi}{2 \beta}$) to restrict the Hilbert space; the complex chemical potential ensures that the partition function contributions from the two unphysical sectors perfectly cancel each other. We note that we define the Green's function (along with their diagrammatic propagators) for the conduction electron and localized $f$ electron in Fig. \ref{fig_greens_fun}.
\begin{figure}[h]
\includegraphics{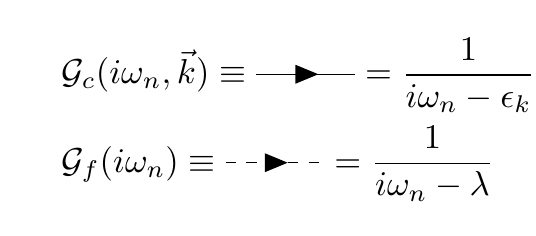}
\caption{Green's function for conduction (solid line) and $f$ (dashed line) electrons.}
\label{fig_greens_fun}
\end{figure}

We are now at the stage to compute the four-point correlation function ($\Gamma$). We present the four point function up to two-loop (or third order in the `Kondo' interaction strength) in Fig. \ref{fig_methods_rg_diag}, where we have organized the diagrams into $\Gamma_{i}$  ($\Gamma_{r}$) to categorize diagrams that are 2-particle irreducible (reducible). 
We compute the vertex function in the limit of $\nu << D$ (where $\nu$ is the energy sum of the incoming conduction and $f$ electrons) and retain all the leading logarithmically divergent terms. It is the irreducible diagrams that provide the logarithmically divergent terms. 
Finally, we apply the Wilsonian RG procedure: the {low-energy} scattering amplitude/rate is independent of the UV cutoff (D) i.e.
$\rho \Gamma(\nu, D, \vec{g}) = \rho  \Gamma(\nu, D',\vec{g}(D'))$,
where $\vec{g}$ and $\vec{g}(D')$ symbolically represent the bare and renormalized coupling constants. Solving for the renormalized coupling constant yields the desired $\beta$ function.
For completeness, we note that under the RG, symmetry allowed terms of the form $\sigma^{0,1,2,3}_{\alpha \beta} c^{\dag}_{x, \alpha} c_{y, \beta}$ are also generated.
Importantly, these terms only introduce/renormalize hopping amplitudes, and not the multipolar Kondo couplings, and as such do not affect the obtained $\beta$-functions.

\begin{figure*}[h]
\includegraphics[scale=0.87]{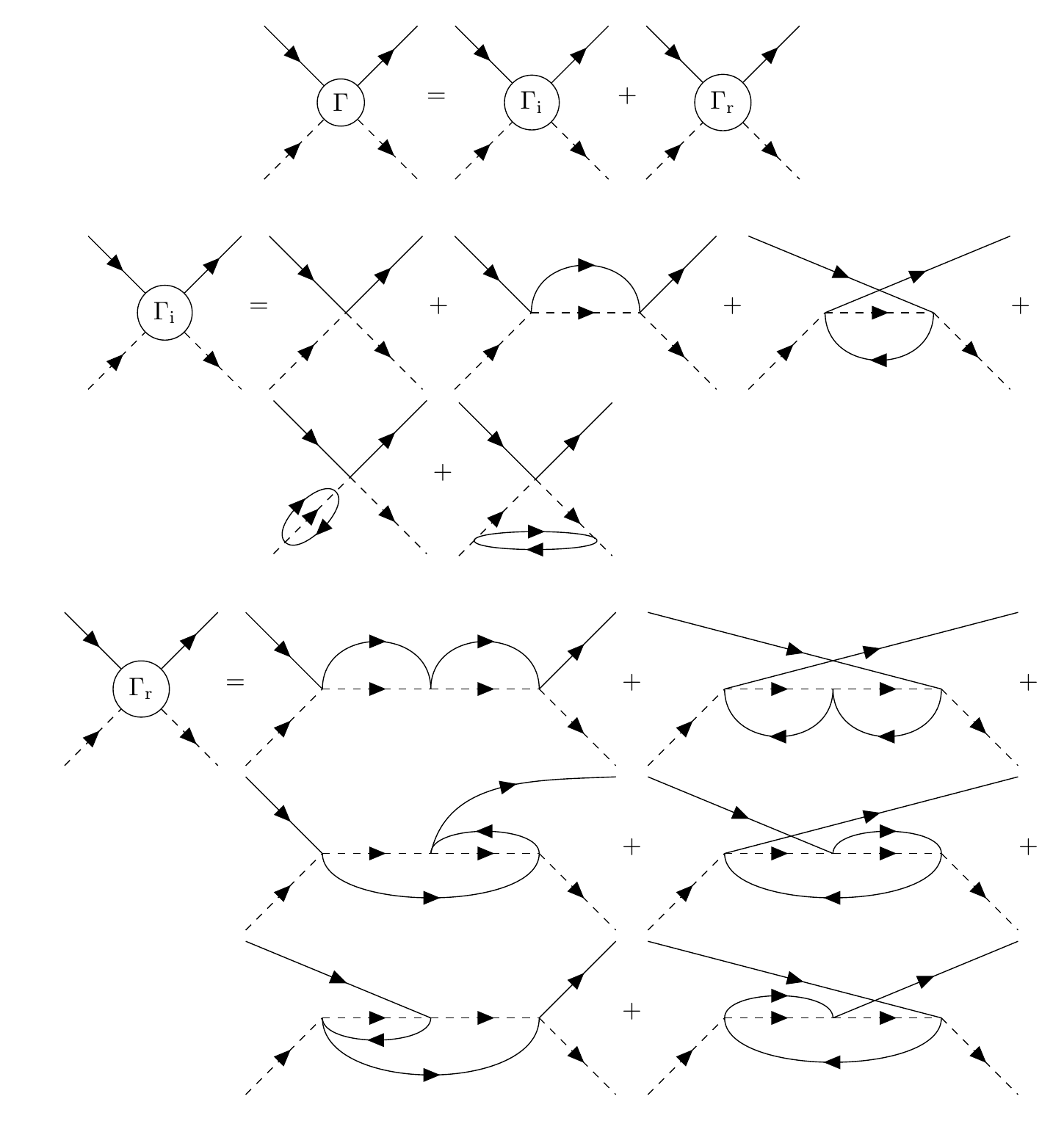}
\caption{Four point function ($\Gamma$) which is composed of 2-particle irreducible ($\Gamma_{\text{i}}$) and reducible ($\Gamma_{\text{r}}$) diagrams. $\Gamma_{\text{i}}$ contains the logarithmically divergent terms necessary for the computation of the $\beta$ function.}
\label{fig_methods_rg_diag}
\end{figure*}

\vfill

\pagebreak
\section{Renormalization group flow diagrams of $e_g$ and $p$ fixed points} \label{si_rg_flow_diag}

In Figs. \ref{fig:Ng1}, \ref{fig:Ng2}, \ref{fig:Ng3}, we depict the RG flow diagrams for the $e_g$ only model, and for the $p$ model (for the plane including the $p_{1 \pm}^*$ and $p_{2 \pm}^*$ fixed points). Due to the similarity of the RG flow diagrams in Figs. \ref{fig:Ng1} and \ref{fig:Ng2} about the respective pair of fixed points, it is suggestive that they describe similar low-energy physics. The RG flow diagram depicted in Fig. \ref{fig:Ng3} is qualitatively distinct in terms of flow nearby to the fixed points.

\begin{figure}[h]
   \includegraphics[width=0.6\linewidth]{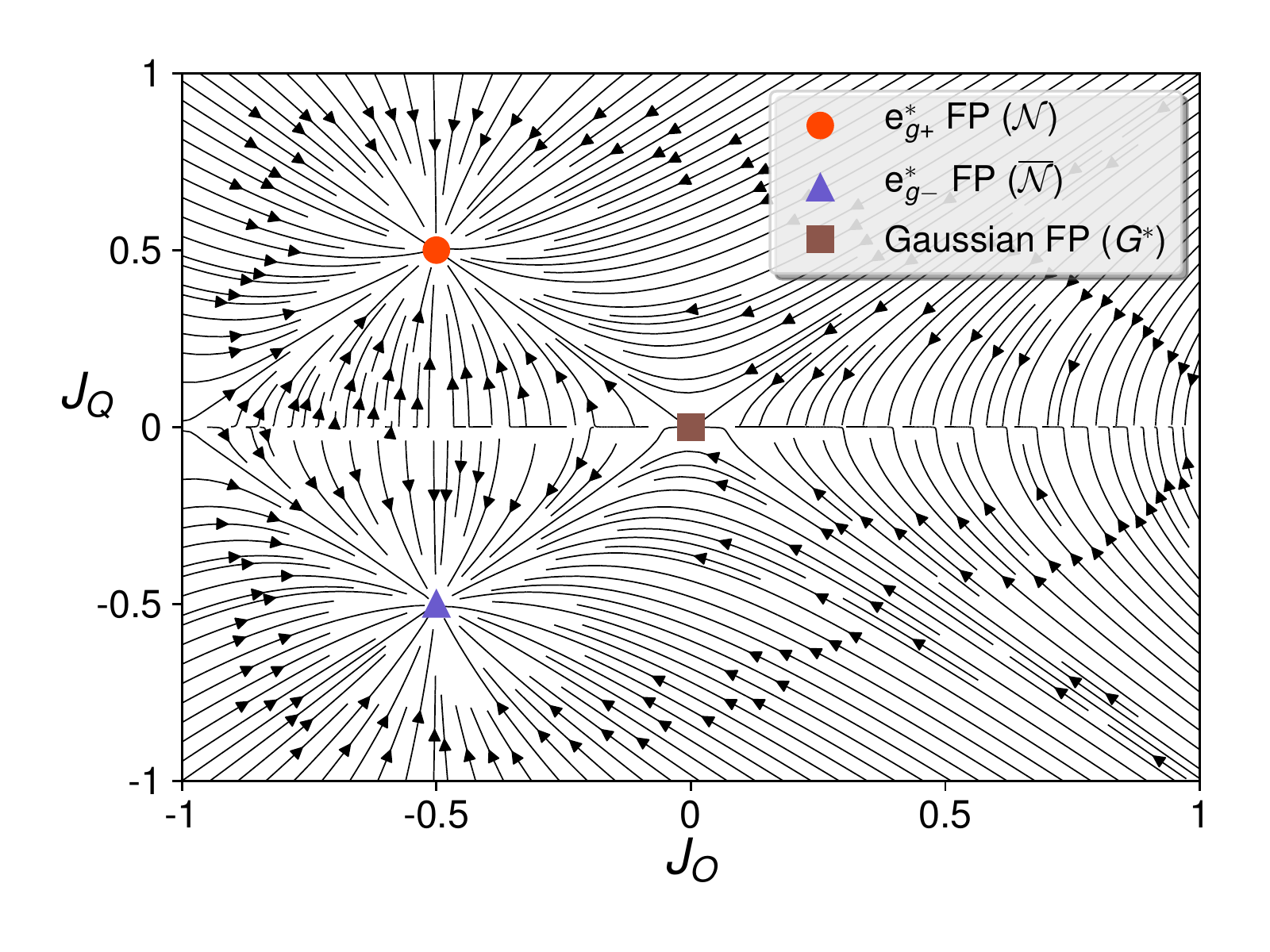}
   \caption{RG flow diagram of $e_g$-only model in the $J_O$-$J_Q$ plane. The Nozi\`eres fixed points ($e_g{\pm}^*$) in Lines $\mathcal{N}$, $\overline{\mathcal{N}}$ are denoted by `red' dot and `blue' triangles, respectively. The Gaussian fixed point is denoted by the `brown' square. The Ising-like line of fixed points, with regions where it is repulsive ($J_O<0$) and attractive ($J_O>0$) is apparent.}
   \label{fig:Ng1} 
\end{figure}

\begin{figure}[h]
   \includegraphics[width=0.6\linewidth]{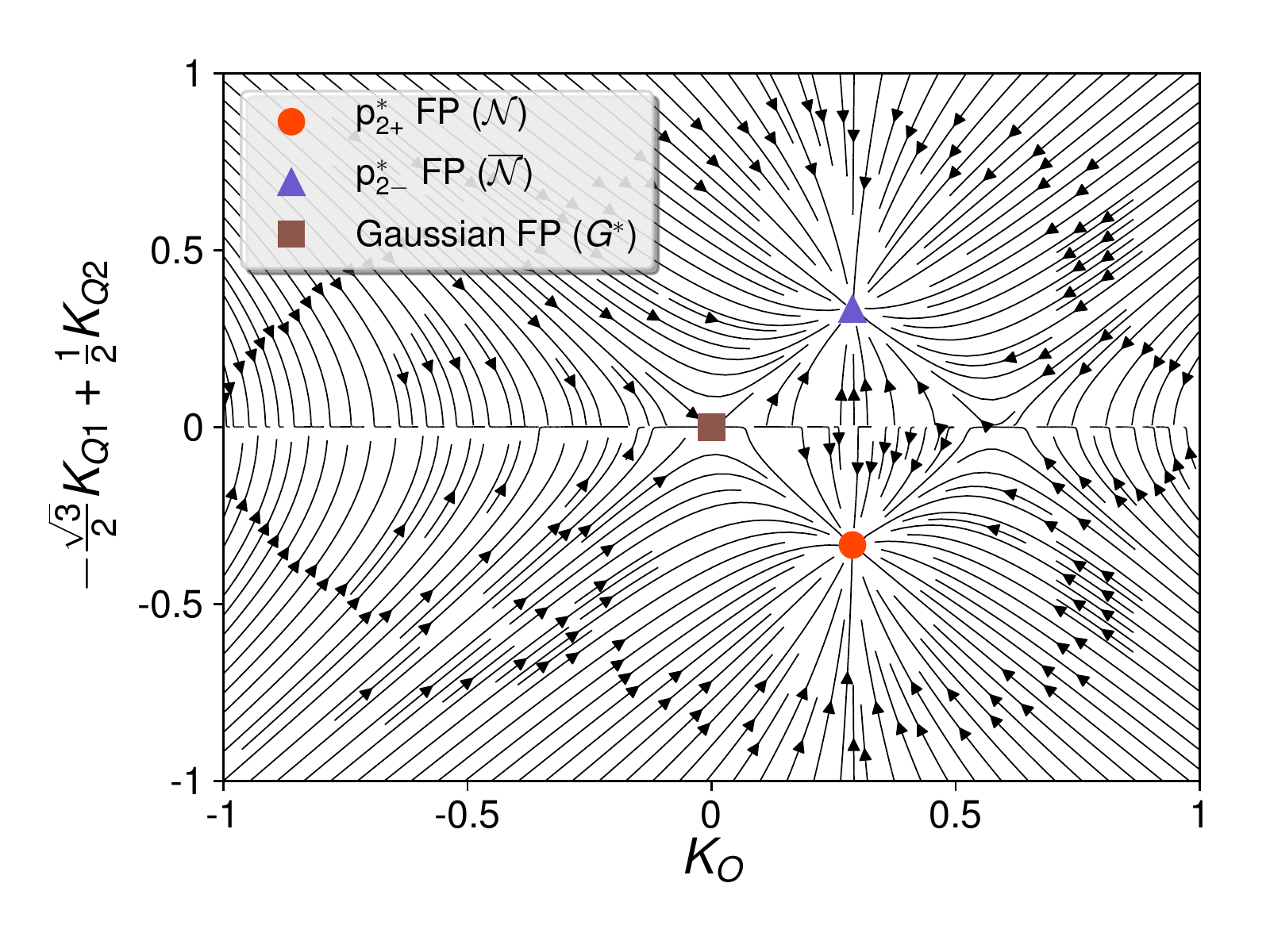}
   \caption{RG flow diagram of $p$-orbital only model in a particular 2D plane spanned by $-\frac{\sqrt{3}}{2}K_{Q1} + \frac{1}{2} K_{Q2}$ and $K_{O}$. The p-only fixed points ($p_{2 \mp}^*$) in Lines $\mathcal{N}$, $\overline{\mathcal{N}}$ are denoted by `red' dot and `blue' triangles, respectively. The Gaussian fixed point is denoted by the `brown' square. The Ising-like line of fixed points is apparent.}
   \label{fig:Ng2}
\end{figure}

\begin{figure}[h]
   \includegraphics[width=0.6\linewidth]{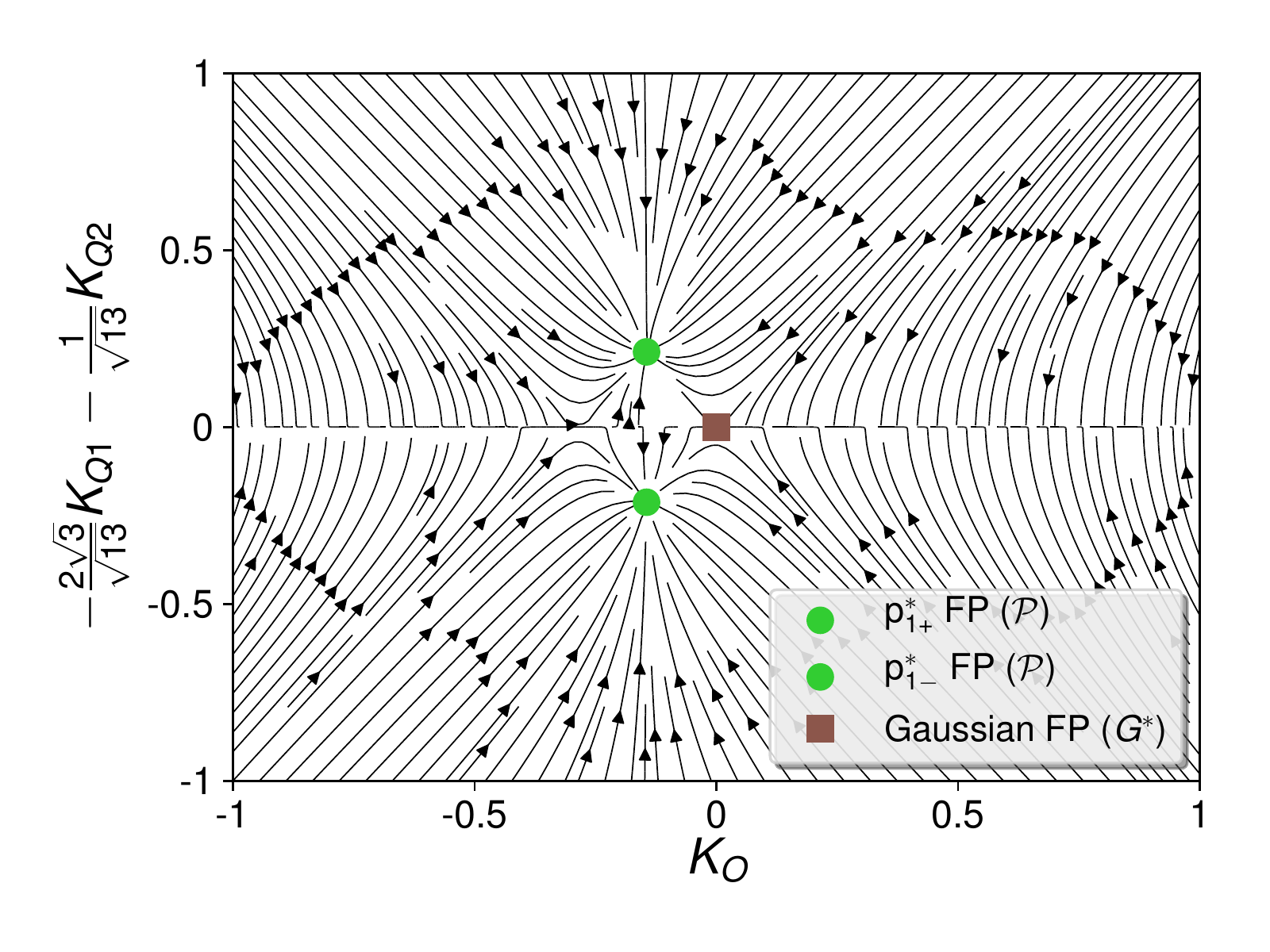}
   \caption{RG flow diagram of $p$-orbital only model in a particular 2D plane spanned by $-\frac{2 \sqrt{3}}{\sqrt{13}}K_{Q1} - \frac{1}{\sqrt{13}} K_{Q2}$ and $K_{O}$. The p-only fixed points ($p_{1 \mp}^*$) in Lines $\mathcal{P}$ are denoted by `green' dots. The Gaussian fixed point is denoted by the `brown' square. The Ising-like line of fixed points is apparent.}
   \label{fig:Ng3}
\end{figure}

\vfill

\section{Hamiltonian and $\beta$ functions: inclusion of $e_g \otimes p$ mixing} \label{app_together_all}

We present here the `Kondo' couplings that mix the $e_g$ and $p$ orbitals. First, we present the `Kondo' Hamiltonians that introduce the three additional couplings: quadrupolar $L_{Q1}, L_{Q2}$ and octupolar $L_O$. 
\begin{align}
{H}_O^{e_g \otimes p} = {L}_O & S^z_j c_{j,a,\alpha}^{\dagger}  \left[  \sigma_{\alpha \beta}^x \otimes \left( -\sqrt{3}  \kappa^{2x,r}_{ab} + \kappa^{1x,r}_{ab} \right)  +   \sigma_{\alpha \beta}^y \otimes \left( \sqrt{3}  \kappa^{2y,r}_{ab} + \kappa^{1y,r}_{ab} \right)  -2 \sigma_{\alpha \beta}^z \otimes \kappa^{1z,r}_{ab}  \right]    c_{j,b,\beta}
\end{align}
\begin{align}
{H}_{Q1}^{e_g \otimes p} = -{L}_{Q1} c_{j,a,\alpha}^{\dagger} & \left[  \left( \frac{\sqrt{3}}{2} S^x_j \right)      \sigma_{\alpha \beta}^x \otimes \kappa^{2x,i}_{ab} + \left( S^x_j - \frac{\sqrt{3}}{2} S^y_j \right) \sigma_{\alpha \beta}^x \otimes \kappa^{1x,i}_{ab}  \right. \nonumber \\ 
& - \left. \left( \frac{\sqrt{3}}{2} S^x _j\right)  \sigma_{\alpha \beta}^y \otimes \kappa^{2y,i}_{ab}   + \left( S^x_j + \frac{\sqrt{3}}{2} S^y_j \right) \sigma_{\alpha \beta}^y \otimes \kappa^{1y,i}_{ab}   \right. \nonumber \\ 
& - \left.  \left( \frac{{3}}{2} S^y_j \right)   \sigma_{\alpha \beta}^z \otimes \kappa^{2z,i}_{ab} - \left( \frac{{S^x_j}}{2}  \right) \sigma_{\alpha \beta}^z \otimes \kappa^{1z,i}_{ab}   \right]  c_{j,b,\beta}
\end{align}
\begin{align}
{H}_{Q2}^{e_g \otimes p} = -{L}_{Q2} c_{j,a,\alpha}^{\dagger} & \left[  \left( \frac{\sqrt{3}}{2} S^x_j + S^y_j \right)       \sigma_{\alpha \beta}^x \otimes \kappa^{2x,i}_{ab}   -  \left( \frac{\sqrt{3}}{2} S^y_j \right)  \sigma_{\alpha \beta}^x \otimes \kappa^{1x,i}_{ab}  \right. \nonumber \\ 
& + \left. \left( -\frac{\sqrt{3}}{2} S^x_j + S^y_j \right)   \sigma_{\alpha \beta}^y \otimes \kappa^{2y,i}_{ab}  +  \left( \frac{\sqrt{3}}{2} S^y_j \right)  \sigma_{\alpha \beta}^y \otimes \kappa^{1y,i}_{ab}   \right. \nonumber \\ 
&  \left. - \left( \frac{{S^y_j}}{2}  \right)     \sigma_{\alpha \beta}^z \otimes \kappa^{2z,i}_{ab} -  \left( \frac{{3}}{2} S^x_j  \right) \sigma_{\alpha \beta}^z \otimes \kappa^{1z,i}_{ab}      \right]  c_{j,b,\beta}
\end{align}
where $\kappa$ are the generalized SU(5) Gell-Mann Matrices describing the orbital degree of freedom $ \left\{ x^2-y^2, 2z^2 - x^2 - y^2 \right \} \otimes \left\{ x,y,z \right \}$ as listed in 
SI \ref{app_pauli_extended}.

The resulting complete RG flow equations are,
\begin{align}
\frac{d K_{Q1}}{d \ln D} &= 6 K_{Q2} K_{O} - \sqrt{3} L_O \left(L_{Q1} + 3 L_{Q2} \right)   \\
&+  K_{Q1} \left( 12 K_{Q2}^2 + 6 K_{O}^2 + 2 K_{Q1}^2 + 24 L_O ^2 + 2 J_Q ^2 + \frac{15}{2} L_{Q1}^2 + 9 L_{Q1} L_{Q2} + \frac{15}{2} L_{Q2}^2 + 2J_O^2 \right) \nonumber \\ \nonumber \\
\frac{d K_{Q2}}{d \ln D} &= K_{O} \left(K_{Q1} - \sqrt{3} K_{Q2} \right) + \frac{L_O}{2} \left(5 L_{Q1} + 3 L_{Q2} \right)   \\
&+  K_{Q2} \left( 12 K_{Q2}^2 + 6 K_{O}^2 + 2 K_{Q1}^2 + 24 L_O ^2 + 2 J_Q ^2 + \frac{15}{2} L_{Q1}^2 + 9 L_{Q1} L_{Q2} + \frac{15}{2} L_{Q2}^2 + 2J_O^2 \right) \nonumber \\  \nonumber \\
\frac{d K_{O}}{d \ln D} &= 4 K_{Q1} K_{Q2}  - 2 \sqrt{3} K_{Q2}^2 - \sqrt{3} \left(L_{Q1} + L_{Q2} \right)^2   \\
&+  K_{O} \left( 24 K_{Q2}^2 + 4 K_{Q1}^2 + 4 J_Q ^2 + {15} L_{Q1}^2 + 18 L_{Q1} L_{Q2} + {15} L_{Q2}^2 \right) \nonumber \\  \nonumber \\
\frac{d J_{Q}}{d \ln D} &= 2 J_Q J_O - 6L_O\left(L_{Q1} + L_{Q2} \right)   \\
&+  J_{Q} \left( 12 K_{Q2}^2 + 6 K_{O}^2 + 2 K_{Q1}^2 + 24 L_O ^2 + 2 J_Q ^2 + \frac{15}{2} L_{Q1}^2 + 9 L_{Q1} L_{Q2} + \frac{15}{2} L_{Q2}^2 + 2J_O^2 \right) \nonumber \\  \nonumber \\
\frac{d J_{O}}{d \ln D} &= 2 J_Q^2 + \frac{9}{4} \left(L_{Q1}^2 + L_{Q2}^2 \right) + \frac{15}{2} L_{Q1} L_{Q2}  \\
&+  J_{O} \left( 24 K_{Q2}^2 + 4 K_{Q1}^2 + 4 J_{Q}^2 + {15} L_{Q1}^2 + 18 L_{Q1} L_{Q2} + {15} L_{Q2}^2  \right) \nonumber \\  \nonumber \\
\frac{d L_{O}}{d \ln D} &= -J_Q \left( L_{Q1} + L_{Q2} \right) -  \frac{K_{Q1}}{2 \sqrt{3}} \left(L_{Q1} + 3 L_{Q2} \right) + \frac{K_{Q2}}{2}  \left( 5 L_{Q1} + 3 L_{Q2} \right)  \\
&+  L_{O} \left( 24 K_{Q2}^2 + 4 K_{Q1}^2 + 4 J_{Q}^2 + {15} L_{Q1}^2 + 18 L_{Q1} L_{Q2} + {15} L_{Q2}^2  \right) \nonumber \\  \nonumber \\
\frac{d L_{Q1}}{d \ln D} &= L_{Q2} J_O - \frac{\sqrt{3}}{2} K_O \left(L_{Q1} + L_{Q2} \right) - L_O \left(J_Q - \frac{K_{Q1}}{\sqrt{3}} - 4 K_{Q2} \right) \\
&+  L_{Q1} \left( 12 K_{Q2}^2 + 6 K_{O}^2 + 2 K_{Q1}^2 + 24 L_O ^2 + 2 J_Q ^2 + \frac{15}{2} L_{Q1}^2 + 9 L_{Q1} L_{Q2} + \frac{15}{2} L_{Q2}^2 + 2J_O^2 \right) \nonumber \\  \nonumber \\
\frac{d L_{Q2}}{d \ln D} &= L_{Q1} J_O - \frac{\sqrt{3}}{2} K_O \left(L_{Q1} + L_{Q2} \right) - L_O \left(J_Q + \sqrt{3} {K_{Q1}} \right) \\
&+  L_{Q2} \left( 12 K_{Q2}^2 + 6 K_{O}^2 + 2 K_{Q1}^2 + 24 L_O ^2 + 2 J_Q ^2 + \frac{15}{2} L_{Q1}^2 + 9 L_{Q1} L_{Q2} + \frac{15}{2} L_{Q2}^2 + 2J_O^2 \right) \nonumber 
\end{align}
The stable lines of fixed points are,
\begin{align}
\text{Line } \mathcal{N}: \nonumber \\
& K_{Q1} = \frac{1 - 2 J_Q}{2 \sqrt{3}}, ~~~~~ K_O = K_{Q1}, ~~~~~  K_{Q2} = -\frac{K_{Q1}}{\sqrt{3}} \\ 
& J_Q \in [0, 1/2] ,  \ \ \ \ \  J_O = - J_Q \nonumber \\ 
& L_O = L_{Q1} = L_{Q2} = \pm \frac{ \sqrt{J_Q(1-2J_Q)} }{2 \sqrt{3}} \nonumber 
\end{align}

\begin{align}
\text{Line } \overline{\mathcal{N}}: \nonumber \\
& K_{Q1} = - \frac{1 + 2 J_Q}{2 \sqrt{3}}, ~~~~~ K_O = -K_{Q1}, ~~~~~  K_{Q2} = -\frac{K_{Q1}}{\sqrt{3}} \\ 
& J_Q \in [-1/2, 0] ,  \ \ \ \ \  J_O =  J_Q \nonumber \\ 
& L_O = - L_{Q1} = - L_{Q2} = \pm \frac{ \sqrt{-J_Q(1+2J_Q)} }{2 \sqrt{3}} \nonumber 
\end{align}

\begin{align}
\text{Line } \mathcal{P}: \nonumber \\
& K_{Q1}  \in \left[-\frac{1}{2 \sqrt{6}}, \frac{1}{2 \sqrt{6}} \right] , ~~~~~ K_O = -2\sqrt{3}K_{Q1}^2, ~~~~~  K_{Q2} = \frac{K_{Q1}}{\sqrt{12}} \\ 
& J_Q = 0 ,  \ \ \ \ \ J_O = \frac{1}{4} - 6 K_{Q1}^2 \nonumber \\ 
& L_O = \pm \frac{ K_{Q1} \sqrt{1 - 24K_{Q1}^2} }{2},  ~~~~~ L_{Q1} = -L_{Q2} =  \mp \frac{ \sqrt{1 - 24K_{Q1}^2} }{4 \sqrt{3}} \nonumber 
\end{align}

\section{Running Coupling constant} \label{method_running}

In studying the physical observables associated with the fixed points, it is necessary to obtain the behaviour of the coupling constant in the vicinity of the fixed point. To do so, we consider general $\beta$ functions associated with the multiple general coupling constants, $\frac{d}{d \ln D} g_l = \beta_l \left(g \right)$,
where $l = 1,2,...$ indexes the multiple couplings. In the vicinity of the fixed point, we have $g_l \approx g_l^*$, where $g_l^*$ is the $l^{th}$ coupling constant's fixed point value. We expand the above general $\beta$ function in the neighbourhood of the fixed point to obtain \cite{weinberg_1996_vol_2},
\begin{align}
\frac{d}{d \ln D} \left( g_l - g_l ^*\right) &= \beta_l \left(g - g^* \right) \approx \beta_l (g^*) + \sum_k \mathbb{M}_k ^l \left( g_k - g_k ^*\right)
\label{eq_beta_diff_general}
\end{align}
where $\mathbb{M}_k ^l=\frac{ \partial \beta_l}{\partial g_k}$ is the Jacobian matrix associated with the $\beta$ function. The first term on the right-hand-side is zero, by definition of a fixed point, and we solve the subsequent matrix differential equation to obtain,
\begin{align}
g_l (D) = g_l^* + \sum_k c_k D^{\lambda_k} u_l^k
\end{align}
where $c_k$ are to-be-determined coefficients, and $\lambda_k$ and $u_l^k$ are the eigenvalues and eigenvectors, respectively, of the Jacobian matrix. Since we have obtained such an expression by examining behaviour in close vicinity of the fixed point, and since the RG flow is physically controlled by tuning down the temperature, we can take (at least dimensionally) $D \sim T << 1$. Doing so, we obtain the temperature-dependent behaviour of the coupling constant -- also known as the \textit{running coupling constant} -- at very low temperatures (and close to the fixed point). As mentioned above, the coefficients $c_k$ are determined by initial conditions of the differential equation \ref{eq_beta_diff_general}, but we can go further by taking all coefficients with $\lambda_k \leq 0$ to be zero. The physical justification is that as $T \rightarrow 0$, we want the running coupling to arrive at the fixed point; if $\lambda_k<0 \implies T^{\lambda_k} \rightarrow \infty$, which corresponds to flowing away from the fixed point, and $\lambda_k=0 \implies T^{\lambda_k} \rightarrow 1$, which entails that (at this order of perturbation), the running coupling does not move from its original location in the parameter space.

\section{Free energy and Specific Heat} \label{method_cv}

To calculate the specific heat, we first need to compute the free energy of the system. We do so by using the linked-cluster expansion theorem,
\begin{align}
F = F_0 - \sum_{l=1}^{\infty} \frac{1}{l} F_l
\end{align}
where $F_0$ is the free energy of the non-interacting (i.e. vanishing `Kondo' interaction) model, and $F_l$ contains all the different, connected diagrams with $l$ `Kondo' interactions. 
Diagrammatically, we can represent each term in the free energy as in Fig. \ref{fig_free_energy}.
\begin{figure}[t]
\includegraphics[scale=1.0]{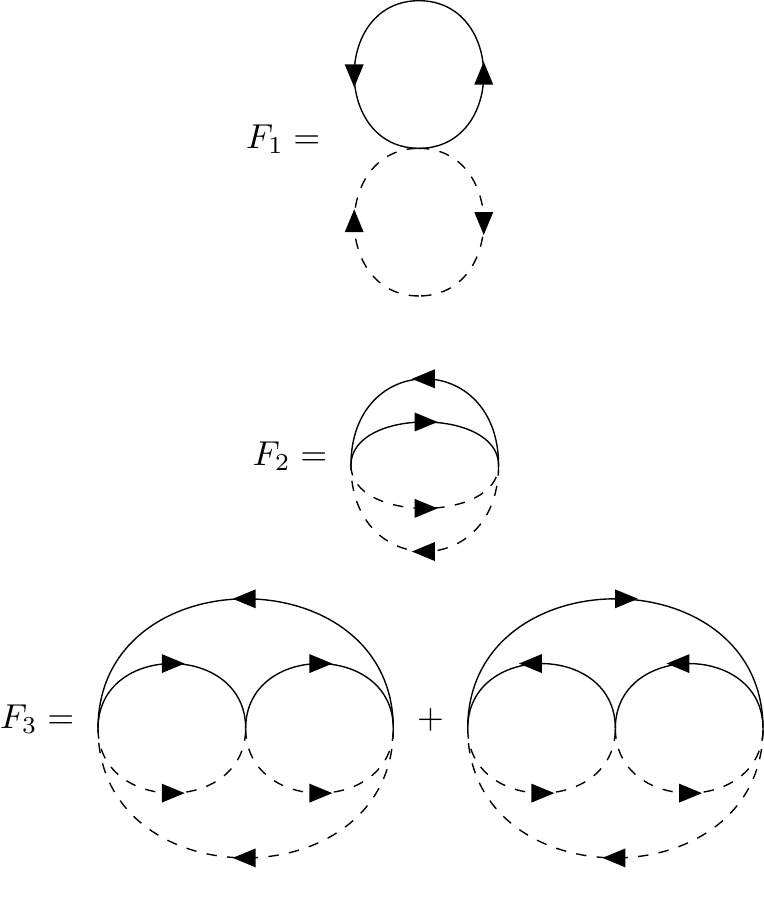}
\caption{Free Energy Diagrams up to third order in perturbation theory.}
\label{fig_free_energy}
\end{figure}
We note that the two diagrams of $F_3$ yield the same expression after computing the Matsubara sums and internal integrals. $F_1$ vanishes due to the internal sum over the pseudospin of a single closed loop is equivalent of taking Tr$[S^i]=0$. To extract out the linear in $T$ behaviour we need to carefully perform the Matsubara sums and energy integrals, as described below. The perturbative evaluation of the free energy is performed using the bare couplings. Once computed, we then apply the RG procedure of taking the bare coupling constants to the running coupling constants. The specific heat is thus finally obtained by taking,
\begin{align}
c_v = - T \frac{ \partial ^2 F} {\partial T^2}
\end{align}

\subsection{Integrals of Free energy} \label{si_free_energy_int}

{In order to obtain the specific heat, it is important to extract the temperature dependence in the free energy. We have both an implicit- and explicit-temperature dependence. The implicit-temperature dependence arises from the running coupling constants, as discussed above. For the models considered in this work, the explicit temperature dependence is identical, regardless of which fixed point/line we consider. Here, we provide details on obtaining the explicit temperature dependence at second and third order in perturbation theory.}

First, we compute the second-order free energy diagrams. Performing the internal summation over the orbital and spin degrees of freedom and the Matsubara summations in the order of fermionic frequencies $i \omega_3$, $ i \omega_2$ and then $ i \nu_1$; this specific order avoids complications regarding the multiplicity of the Matsubara pole structure. For the second order diagram, we obtain the following expression,
\begin{align}
F_2 =  -\frac{1}{2} \mathcal{F}_2 \int_{-D} ^{D} d \epsilon_1 d \epsilon_2 \frac{f(\epsilon_1) - f(\epsilon_2)}{\epsilon_1 - \epsilon_2},
\end{align}
where $\mathcal{F}_2 = (4 J_Q^2+24 K_{Q2}^2+6 K_O^2+4 K_{Q1}^2+24 L_O^2+2 J_O^2+18 L_{Q1} L_{Q2}+15 L_{Q1}^2+15 L_{Q2}^2)$. {The implicit-temperature dependence mentioned above is encoded in $\mathcal{F}_2$.} 
We compute this integral by first performing the integral over $\epsilon_1$, and then we perform a Sommerfeld expansion. Specifically, we avoid the vanishing denominator by choosing the domains of $\epsilon_1$ and $\epsilon_2$ symmetrically about the problematic domain of integration i.e. when $\epsilon_1 = \epsilon_2$. The subsequent {explicit-temperature} result is a constant + $T^2$ scaling behaviour, as is commonly seen in Sommerfeld expansions.

Second, we consider the third-order free energy correction. The Matsubara summations are again performed in the order of $i \omega_3$, $ i \omega_2$ and then $ i \nu_1, i \nu_2$, to avoid the same multiplicity of pole issue described in $F_2$ above. After lengthy simplifications, we arrive at the following expression of,
\begin{align}
F_3 = \frac{1}{2} \mathcal{F}_3 \int_{-D} ^{D} d \epsilon_1 d \epsilon_2 d \epsilon_3 \Bigg( \frac{f(-\epsilon_1) f(\epsilon_2) f(\epsilon_3)}{(\epsilon_1 - \epsilon_2)(\epsilon_1 - \epsilon_3)} + \frac{f(\epsilon_1) f(-\epsilon_2) f(\epsilon_3)}{(\epsilon_2 - \epsilon_1)(\epsilon_2 - \epsilon_3)} + \frac{f(\epsilon_1) f(\epsilon_2) f-\epsilon_3)}{(\epsilon_3 - \epsilon_1)(\epsilon_3 - \epsilon_2)} \Bigg),
\end{align}
where $\mathcal{F}_3 = \frac{3}{4} [48 J_Q L_O(L_{Q1} + L_{Q2} ) - 8 J_Q^2 J_O - 24 K_{Q2} \left(2 K_O K_{Q1} + L_O \left(5 L_{Q1} + 3 L_{Q2} \right)\right)+24 \sqrt{3} K_{Q2} ^2 K_O+ \\ 8 \sqrt{3} K_{Q1} L_O \left(L_{Q1} + 3 L_{Q2} \right)+24 \sqrt{3} K_O L_{Q1} L_{Q2} + 12 \sqrt{3} K_O (L_{Q1}^2 + L_{Q2}^2)-30 J_O L_{Q1} L_{Q2} - 9 J_O (L_{Q1}^2 + L_{Q2}^2) ]$. {The implicit-temperature dependence mentioned above is encoded in $\mathcal{F}_3$.} To extract out the {explicit} temperature dependence from this term, we implement the procedure performed in Refs. \onlinecite{kondo_linear_t_free_energy, Gan_1994}. The key issue is that although this term on the whole is convergent, (computing one at a time) the individual pieces are not. To circumvent this issue, we consider taking the Cauchy principal value of the integral (i.e. $I_3 \equiv F_3 |_{\delta}$) and then we compute the difference between the original expression and this principal-valued integral. Formally, this is performed by taking,
\begin{align}
\frac{1}{x} \rightarrow \frac{1}{(x)_\delta} = \lim_{\delta \rightarrow 0} \frac{1}{x^2 + \delta^2} 
\label{eq_delta_helpful}
\end{align}
The rest of the calculation is analogous to that presented in Ref. \onlinecite{Gan_1994}. We indicate that it is helpful to perform the above $\delta$-integrals by using $(x^2 + \delta^2) = (x + i\delta)(x - i\delta)$, and carefully performing the contour integration. The final result, after lengthy algebra, is {(for the explicit-temperature dependence)} a const. + $T$ scaling behaviour (from both the principal value integral and the `difference' terms), where the constant is the zero-temperature result.

\section{Self Energy and Resistivity} \label{method_resistivity}

\begin{figure}[t]
\includegraphics[scale=1.0]{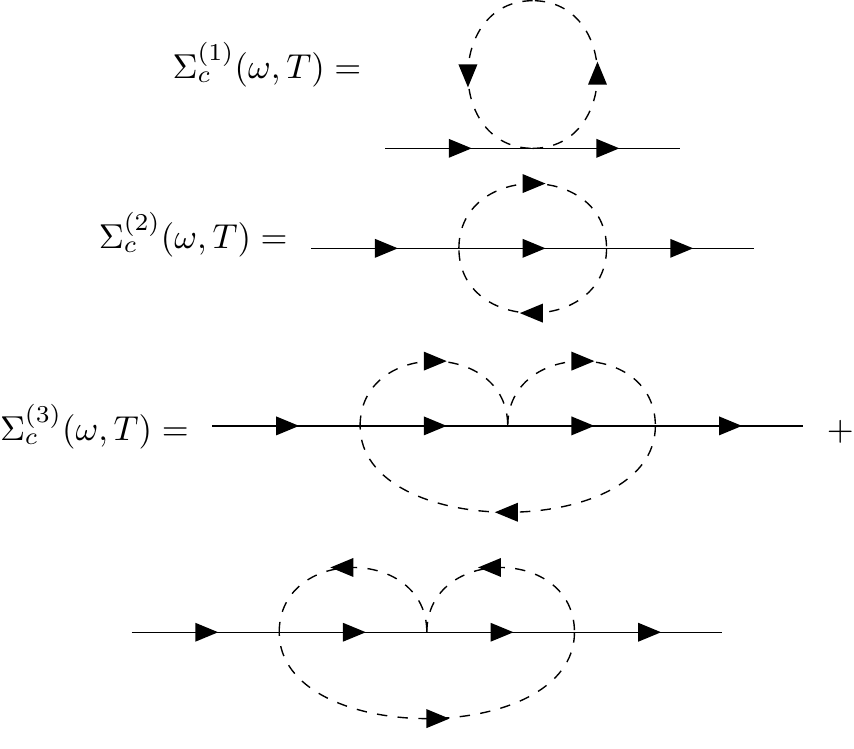}
\caption{Self Energy Diagrams up to third order in perturbation theory.}
\label{fig_self_energy}
\end{figure}

The electrical resistivity is intimately related to the total relaxation lifetime (or scattering rate) of the conduction electron, which is taken to be
\begin{align}
\frac{1}{\tau_{\text{tot}}} = \frac{1}{\tau_0} + \frac{1}{\tau_K} \ ,
\end{align}
where $\tau_{\text{tot}}$ is the total conduction electron relaxation lifetime (or total time between scattering events), $\tau_0$ is the conduction electron lifetime in the absence of the `Kondo' interaction. $\tau_K$ is the relaxation lifetime correction arising from the `Kondo' interaction with the multipolar impurity. As is typical in the single impurity Kondo problem, the ordinary scattering rate in the absence of Kondo-scattering impurities ($1/\tau_0$) is taken to be much larger than the scattering rate due to the Kondo impurity due to the sheer scarcity of the number of multipolar impurities, i.e. $\tau_0 \ll \tau_K$. $\tau_K$ arises via the Kondo exchange and is obtained from
\begin{align}
\frac{1}{ \tau_{K} (\omega, T)} = - 2 \operatorname{Im} \left(\Sigma_c (\omega, T)\right) \ ,
\label{equ_tau_K}
\end{align}
where $\Sigma_c (\omega, T)$ is the conduction electron self-energy. Diagrammatically, the self energy can be represented as sum of the  Feynman graphs in Fig. \ref{fig_self_energy}.

The self energy diagrams share a similarity with the free energy diagrams; this is not a coincidence as the free energy diagrams can be thought of as the self energy diagrams but with the external conduction legs joined together. Just as in the free energy, the first order diagram vanishes due to taking Tr$[S^i]=0$. We note that due to the matrix structure associated with the multipolar Kondo interaction, the self-energy possesses a $10 \times 10$ matrix structure for the full model. We negate this difficulty by taking $ \Sigma_c (\omega, T)$ in Eq. \ref{equ_tau_K} to be the sum of the eigenvalues of the self energy matrix. 

The DC conductivity can subsequently be obtained by insertion of the computed self energy into the standard conductivity expression \cite{abrikosov_2_65},
\begin{align}
 \sigma (T) = \frac{n_e e^2}{m_e} \int_0 ^{\infty} d \omega \frac{\tau_{\text{tot}} (\omega)}{2T \cosh^2(\omega/2T)},
\label{eq_abrikosov_conduct}
\end{align}
where $n_e$, $e$ and $m_e$ are the conduction electrons' density, charge and mass, respectively, and $T$ is the temperature. We expand the total relaxation time in Eq. \ref{eq_abrikosov_conduct} as $\tau_{\text{tot}}(\omega) \approx \tau_0 \left(1 - \frac{\tau_0}{\tau_K(\omega)} \right)$, and invert Eq. \ref{eq_abrikosov_conduct} expression to yield the resistivity $\rho \equiv \rho(T) - \rho_0$, where $\rho_0 = \frac{m_e}{n_e e^2}$ is the ordinary resistivity and $\rho$ is the contribution arising from the multipolar Kondo exchange. Finally, we apply the RG procedure of taking the bare coupling constants to the running coupling constants to obtain the low-temperature behaviour of the resistivity.

\twocolumngrid


%

\end{document}